\documentclass[a4paper]{article}

\usepackage{a4wide}
\usepackage[UKenglish]{babel}
\usepackage{graphicx,subfigure}
    \graphicspath{{./}{figure/}}
\usepackage[colorlinks=true,linkcolor=blue,citecolor=red]{hyperref}
\usepackage{amsmath,amsthm,mathtools,bbold}

\DeclarePairedDelimiter\abs{\lvert}{\rvert}
\DeclarePairedDelimiter\ave{\langle}{\rangle}
\newcommand{\e}{\epsilon}
\newcommand{\ff}{\hat{f}}
\newcommand{\N}{\mathbb{N}}
\renewcommand{\P}{\mathbb{P}}
\newcommand{\pr}[1]{{}^\prime\!#1}
\newcommand{\R}{\mathbb{R}}
\newcommand{\Z}{\mathbb{Z}}

\theoremstyle{remark}\newtheorem{remark}{Remark}

\begin{document}
\title{Multiple-interaction kinetic modelling of a virtual-item gambling economy}

\author{Giuseppe Toscani\thanks{Department of Mathematics ``F. Casorati'', University of Pavia, Via Ferrata 1, 27100 Pavia, Italy
            (\texttt{giuseppe.toscani@unipv.it})} \and
        Andrea Tosin\thanks{Department of Mathematical Sciences ``G. L. Lagrange'', Politecnico di Torino, Corso Duca degli Abruzzi 24, 10129 Torino, Italy
            (\texttt{andrea.tosin@polito.it})} \and
        Mattia Zanella\thanks{Department of Mathematical Sciences ``G. L. Lagrange'', Politecnico di Torino, Corso Duca degli Abruzzi 24, 10129 Torino, Italy
            (\texttt{mattia.zanella@polito.it})}}
\date{}

\maketitle

\begin{abstract}
In recent years, there has been a proliferation of online gambling sites, which made gambling more accessible with a consequent rise in related problems, such as addiction. Hence, the analysis of the gambling behaviour at both the individual and the aggregate levels has become the object of several investigations. In this paper, resorting to classical methods of the kinetic theory, we describe the behaviour of a multi-agent system of gamblers participating in lottery-type games on a virtual-item gambling market. The comparison with previous, often empirical, results highlights the ability of the kinetic approach to explain how the simple microscopic rules of a gambling-type game produce complex collective trends, which might be difficult to interpret precisely by looking only at the available data.

\medskip


\end{abstract}

\section{Introduction}
Gambling is usually perceived as a complex multi-dimensional activity fostered by several different motivations~\cite{Binde}. Due to the rapid technological developments, in the last decade the possibility of online gambling has enormously increased~\cite{Kristiansen}, leading to the simultaneous rise of related behavioural problems. As remarked in~\cite{Jonsson2017}, structural characteristics of online gambling, such as the speed and the availability, led to conclude that online gambling has a high potential risk of addiction.

A non-secondary aspect of the impressive increase in online gambling sites is related to economic interests. Indeed, the expansion of the video-gaming industry has resulted in the formation of a new market, in which gamblers are the actors, that has reached a level of billions of dollars. The continuous expansion of this market depends on many well-established reasons, which include its easy accessibility, low entry barriers and immediate outcome.

As documented in~\cite{Wang}, mathematical modelling of these relatively new phenomena attracted the interest of current research, with the aim of understanding the aggregate behaviour of a system of gamblers. In~\cite{Wang}, the behaviour of online gamblers has been studied by methods of statistical physics. In particular, the analysis has been focused on a popular type of virtual-item gambling, the jackpot, i.e. a lottery-type game which occupies a big portion of the gambling market on the web. As pointed out in~\cite{Wang}, to be able to model the complex online gambling behaviour at both the individual and the aggregate levels is quickly becoming a pressing need for adolescent gambling prevention and eventually for virtual gambling regulation.

The gambling datasets used in~\cite{Wang} have been extracted from the publicly available history page of a gambling site. The huge number of gambling rounds, and the time period (more than seven months) taken into account, allowed for a consistent fitting. The analysis of the dataset has been essentially split in two main parts. A first part deals with the behavioural distribution of the gambler activities. Here, the main result concerns the cumulative distribution function of the number of rounds played by individual gamblers, which was found to be best fitted by a lognormal distribution. A second part of the analysis in~\cite{Wang} is concerned instead with the study of the distribution function of the winnings and of the related correlations. As it happens in many socio-economic phenomena involving multi-agent systems~\cite{pareschi2013BOOK}, the best fitting curve for the winnings has been found to be a power-law-type distribution with cut-off. While the possible reasons leading to the formation of a lognormal distribution for the number of rounds played by the gamblers has been left largely unexplored, the formation of a power law distribution for the incomes has been explained in~\cite{Wang} by resorting to three different random walk models. As clearly outlined by the authors, their aim was to gain insights into the ingredients necessary to obtain from these models results with qualitative properties similar to those of the data derived from the gambling logs.

The huge number of gamblers and the well-defined rules of the game allow us to treat the system of gamblers as a particular multi-agent economic system, in which the agents invest (risk) part of their personal wealth to obtain a marked improvement of their economic conditions. Unlike classical models of the trading activity~\cite{pareschi2013BOOK}, in this gambling economy particular attention needs to be paid to the behavioural reasons pushing people to gamble even in presence of high risks. By looking at the jackpot game from this perspective, and resorting to the classical modelling of multi-agent systems via kinetic equations of Boltzmann and Fokker-Planck type~\cite{pareschi2013BOOK}, we will be able to obtain a detailed interpretation of the datasets collected in~\cite{Wang}.
 
This approach has proved to be powerful in many situations, ranging from the formation of knowledge in a modern society~\cite{Gua3,Pareschi2014} to the spreading of the popularity of online content~\cite{Tosin} or the description of the reasons behind the formation of a lognormal profile in various human activities characterised by their skewness~\cite{Gua1}.
 
Our forthcoming analysis will be split in two parts. In a first part, we will discuss the kinetic modelling of the jackpot gambling and we will study, in particular, the distribution in time of the tickets played and won by the gamblers. Our modelling approach is largely inspired by the similarities of the jackpot game with the so-called \textit{winner takes it all} game described in detail in~\cite{pareschi2013BOOK}. Nevertheless, the high number of gamblers taking part to the game,  the presence of a percentage cut on the winnings operated by the site,  and  the continuous refilling of tickets to play, introduce essential differences. 

In a second part, we will deal with the behavioural aspects linked to the online gambling. This is a phenomenon that may be fruitfully described by resorting to a skewed distribution and that, consequently, may be modelled along the lines of the recent papers~\cite{Gua1,Gua2}. The behavioural aspects of the gambling and their relationships with other economically relevant phenomena have been discussed in a number of papers, cf. e.g.~\cite{Lund2008} and the references therein. Also, the emergence of the skewed lognormal distribution was noticed before. The novelty of the present approach is that we enlighten the principal behavioural aspects at the basis of a reasonable kinetic description.

Going back to the kinetic description of the jackpot game, it is interesting to remark that some related problems have been studied before. The presence of the site cut, which can be regarded as a sort of dissipation, suggests that the time evolution of the distribution function of the tickets played and won by the gamblers may be described in a way similar to other well-known dissipative kinetic models, such as e.g. that of the Maxwell-type granular gas studied by Ernst and Brito~\cite{Ernst2002} or that of the Pareto tail formation in self-similar solutions of an economy undergoing recession~\cite{Slanina}. However, essential differences remain. Unlike the situations described in~\cite{Ernst2002,Slanina}, where the loss of the energy or of the mean value, respectively, was artificially restored by a suitable scaling of the variables, in the present case the percentage cut on each wager, leading to an exponential loss of the mean value of the winnings, is refilled randomly because of the persistent activity of the gamblers even in the presence of losses. A second difference concerns the necessity to take into account a high number of participants in the jackpot game. In~\cite{Wang}, it is conjectured that the shape of the steady state distribution emerging from the game rules does not change as the number of participants increases. Consequently, all models studied there were limited to describe the evolution of winnings in a game with a very small number of gamblers. Here, we adopt instead a different strategy, inspired by the model introduced by Bobylev and Windfall~\cite{Bob11}. In that paper, it is shown that the kinetic description of an economy with transactions among a huge number of trading agents can be suitably linearised, leading to a simpler description. Hence, following~\cite{Bob11}, we will consistently simplify the jackpot game description by introducing a suitable linearisation of the problem, which makes various explicit computations possible.

Out of the detailed kinetic description of the online jackpot game, and unlike the analysis proposed in~\cite{Wang}, we conclude that the game mechanism does not actually give rise to a power-law-type steady distribution of the tickets played and won by the gamblers. The formation of such a fat tail may, however, be obtained by resorting to a different linearisation of the game, which, while apparently close to the actual non-linear version, may be shown numerically to produce quite different trends. 

In more detail, the paper is organised as follows. In Section~\ref{Boltzmann}, we introduce the microscopic model of the jackpot game with $N$ gamblers and its non-linear Boltzmann-type kinetic description with multiple-interactions (Section~\ref{sect:Maxwell}). Next, in the limit of $N$ large, we replace the $N$-interaction dynamics with a sort of \textit{mean field} individual interaction, which gives rise to a linear Boltzmann-type model (Section~\ref{linea1}). We study the large time trend of the linear model by means of a Fokker-Planck asymptotic analysis, which shows that no fat tails are produced at the equilibrium (Section~\ref{sect:gamma}). Finally, by resorting to a different linearisation of the multiple-interaction model based on the preservation of the first two statistical moments of the distribution function, we produce an alternative kinetic model, whose equilibrium distribution has indeed a power-law-type fat tail (Section~\ref{fat}). Nevertheless, we argue that such a new linear model does not provide a description of the gambling dynamics completely equivalent to the original multiple-interaction model and, hence, that it does not describes exactly the original jackpot game. In Section~\ref{Behavior}, we discuss a model for the distribution of the tickets which the gamblers purchase to participate in successive rounds of the jackpot game. This study complements the previous one on the gambling dynamics, as it provides the basis to model the refilling of tickets mentioned above. In Section~\ref{Numerics}, we illustrate the evolution of the real game predicted by the multiple-interaction kinetic model and that of the various linearised models by means of several numerical experiments, which confirm the theoretical findings of the previous sections. Finally, in Section~\ref{sec:conclusions}, we summarise the main results of the work.

\section{Kinetic models of jackpot games}
\label{Boltzmann}
\subsection{Maxwell-type models}
\label{sect:Maxwell}
The jackpot game we are going to study is very simple to describe. At given intervals of time, which may last from a few seconds to several minutes, the site opens a new round of the game that the gamblers may attend. The gamblers participate in the game by placing a bet with a certain number of lottery tickets purchased with one or several skins deposited to the gambling site. There is only one winning ticket in each round of the game. The winning ticket is drawn when the total number of skins deposited as wagers in that round exceeds a certain threshold. The draw is based on a uniformly distributed random number with a range equal to the total number of tickets purchased in that round. The gambler who holds the winning ticket wins all the wagers, i.e. the deposited skins in that round, after a site cut (percentage cut) has been subtracted.
  
As usual in the kinetic description, we assume that the gamblers are indistinguishable~\cite{pareschi2013BOOK}. This means that, at any time $t\geq 0$, the state of a gambler is completely characterised by their wealth, expressed by the number $x\geq 0$ of owned tickets. Consequently, the microscopic state of the gamblers is fully characterised by the density, or distribution function, $f=f(x,\,t)$.

The precise meaning of the density $f$ is the following. Given a subdomain $D\subseteq\R_+$, the integral
$$ \int_D f(x,\,t)\,dx $$
represents the number of individuals possessing a number $x\in D$ of tickets at time $t\geq 0$. We assume that the density function is normalised to one, i.e.
$$ \int_{\R_+}f(x,\,t)\,dx=1, $$
so that $f$ may be understood as a probability density.

The time evolution of the density $f$ is due to the fact that rounds are programmed at regular time intervals and gamblers continuously upgrade their number of tickets $x$ at each new round. In analogy with the classical kinetic theory of rarefied gases, we refer to a single upgrade of the quantity $x$ as an \textit{interaction}.

The game has evident similarities with the \textit{winner takes it all} game described in detail in~\cite[Chapter 5]{pareschi2013BOOK}. The main differences are the presence of a high number of participants and of the site cut. Indeed, while the microscopic interactions in the \textit{winner takes it all} game are pointwise conservative, any round of the online jackpot game leads to a loss of the value returned to the gamblers.

Let us consider a number $N$ of gamblers, with $N\gg 1$, who participate in a sequence of rounds. At the initial time, the gamblers (indexed by $k=1,\,\dots,\,N$) buy certain numbers $x_k=x_k(0)$ of tickets, with the intention to play for a while. While it is clear that actually $x_k\in\N_+$, in order to avoid inessential difficulties, and without loss of generality, we will consider $x_k\in\R_+$. Moreover, we may fix a unitary price for the tickets, so as to identify straightforwardly the number of tickets with the amount of money owned by the gamblers. We assume that each gambler participates in a round by using only a small fraction of their tickets, say $\e\alpha_k x_k$, where $0<\e\ll 1$ while the $\alpha_k$'s may be either constant or random coefficients. In the simplest case, i.e. $\alpha_k=1$ for all $k$, the total number of tickets played by the gamblers in a single round is $\e\sum_{k=1}^{N}x_k$.

At fixed time intervals of length $\Delta t>0$, a ticket is chosen randomly. The owner of that ticket wins an amount of money corresponding to the value of the total number of tickets played in that round, minus a certain fixed cut operated by the site. Let us denote by $x_k(t-1)$ the number of tickets possessed by the $k$th gambler right before the next round. If $\delta>0$ denotes the percentage cut operated by the site, after the new round the quantities $x_k(t-1)$ update to 
\begin{equation}
     x_k(t)=(1-\e)x_k(t-1)+\e (1-\delta)\sum_{j=1}^{N}x_j(t-1)I(A(t-1)-k), \qquad k=1,\,2,\,\dots,\,N.
    \label{n1}
\end{equation}
In~\eqref{n1}, $A(t-1)\in\{1,\,\dots,\,N\}$ is a discrete random variable giving the index of the winner in the forthcoming round. Since the winner is chosen by extracting uniformly one of the played tickets, the random variable $A(t-1)$ may be characterised by the following law:
\begin{equation}
    \P(A(t-1)=k)=\frac{x_k(t-1)}{\sum\limits_{j=1}^{N} x_j(t-1)}, \qquad k=1,\,2,\,\dots,\,N.
    \label{Akk}
\end{equation}
Furthermore, in~\eqref{n1} the function $I(n)$, for $n\in\Z$, is defined by
$$ I(0)=1, \qquad I(n)=0 \quad \forall\,n\neq 0. $$

Because of the fixed cut operated by the site, the total number of tickets, viz. the amount of money, in the hands of the gamblers diminishes at each round, so that, in the long run, the gamblers remain without tickets to play. On the other hand, as noticed in the recent analysis~\cite{Wang}, the data published by the jackpot site certify that this never happens. One may easily identify at least two explanations. First, gamblers with high losses are continuously replaced by new gamblers entering the game. Second, in presence of repeated losses the gamblers continuously refill the amount of money available to their wagers by drawing on their personal reserves of wealth. Notice that we may easily identify the new gamblers entering the game with those leaving it by simply assuming that the number $N$ of gamblers remains constant in time. Taking this non-secondary aspect into account, we modify the upgrade rule~\eqref{n1} as follows:
\begin{equation}
     x_k(t)=(1-\e)x_k(t-1)+\e\beta Y_k(t-1)+\e(1-\delta)\sum_{j=1}^{N}x_j(t-1)I(A(t-1)-k),
    \label{n-gen}
\end{equation}
$k=1,\,\dots,\,N$. In~\eqref{n-gen}, $\beta\geq 0$ is a fixed constant, which identifies the rate of refilling of the tickets. Moreover, the $Y_k$'s are non-negative, independent and identically distributed random variables giving the number of refilled tickets. In agreement with \cite{Wang}, and as explained in full details in Section \ref{Behavior}, one can reasonably assume that the random variables $Y_k$ are lognormally distributed.

The upgrade rules~\eqref{Akk},~\eqref{n-gen} lead straightforwardly to a Boltzmann-type kinetic model describing the time evolution of the density $f(x,\,t)$ of a population of gamblers who play an $N$-player jackpot game, independently and repeatedly, according to the interaction
\begin{equation}
    x_k'=(1-\e)x_k+\e\beta Y_k+\e(1-\delta)\sum_{j=1}^{N}x_jI(A-k), \qquad k=1,\,2,\,\dots,\,N,
    \label{micro}
\end{equation}
where $A\in\{1,\,\dots,\,N\}$ is a discrete random variable with law
$$ \P(A=k)=\frac{x_k}{\sum\limits_{j=1}^{N}x_j}, \qquad k=1,\,\dots,\,N. $$
In~\eqref{micro}, the quantity $x_k$ represents the number of tickets, hence the amount of money, put into the game by the $k$th gambler, while the quantity $x_k'$ is the new number of tickets owned by the $k$th gambler after the draw of the winning ticket.

Starting from the microscopic interaction~\eqref{micro}, the study of the time evolution of the distribution function $f$ may be obtained by resorting to kinetic collision-like models~\cite{pareschi2013BOOK}. Specifically, the evolution of any \textit{observable quantity} $\varphi$, i.e. any quantity which may be expressed as a function of the microscopic state $x$, is given by the Boltzmann-type equation
\begin{equation}
    \frac{d}{dt}\int_{\R_+}\varphi(x)f(x,\,t)\,dx=
        \frac{1}{\tau N}\int_{\R_+^N}\sum_{k=1}^{N}\ave*{\varphi(x_k')-\varphi(x_k)}\prod_{j=1}^{N}f(x_j,\,t)\,dx_1\,\cdots\,dx_N,
    \label{kine1}
\end{equation}
where $\tau$ denotes a relaxation time and $\ave{\cdot}$ is the average with respect to the distributions of the random variables $Y_k$, $A$ contained in~\eqref{micro}. Note that the interaction term on the right-hand side of~\eqref{kine1} takes into account the whole set of gamblers, and consequently it depends on the $N$-product of the density functions $f(x_j,\,t)$, $j=1,\,\dots,\,N$. Thus, the evolution of $f$ obeys a highly non-linear Boltzmann-type equation.

\begin{remark}
In the classical kinetic theory of rarefied gases, the binary collision integral depends on a non-constant \textit{collision kernel}, which selects the collisions according to the relative velocities of the colliding particles. Conversely, the interaction integral in~\eqref{kine1} has a constant kernel, chosen equal to $1$ without loss of generality. This corresponds, in the jargon of the classical kinetic theory, to consider \textit{Maxwellian interactions}. Remarkably, in the case of the jackpot game, this assumption corresponds perfectly to the description of the game under investigation, since one may realistically assume that the numbers of tickets played by different gamblers are uncorrelated.
\end{remark}

Choosing $\varphi(x)=1$ in~\eqref{kine1} yields
$$ \frac{d}{dt}\int_{\R_+}f(x,\,t)\,dx=0, $$
meaning that the total mass of the system is conserved in time. It is worth pointing out that, as a matter of fact, this is the only conserved quantity in~\eqref{kine1}.

In order to better understand the time evolution of $f$, as well as the role of the site cut, we begin by considering the situation in which the gamblers do not refill their tickets, which corresponds to letting $\beta=0$. In this case, the interactions~\eqref{micro} being linear in the $x_k$'s, we can compute explicitly the evolution in time of the mean number of tickets
$$ m(t):=\int_{\R_+}xf(x,\,t)\,dx $$
owned by the gamblers. Indeed, since
\begin{equation}
    \ave*{\sum_{k=1}^{N}x_k'}=(1-\e)\sum_{k=1}^{N}x_k+\e(1-\delta)\sum_{j=1}^{N}x_j\sum_{k=1}^{N}\P(A=k)
        =(1-\e\delta)\sum_{k=1}^{N}x_k,
    \label{mean}
\end{equation}
choosing $\varphi(x)=x$ in~\eqref{kine1} we obtain
\begin{equation}
    \frac{dm}{dt}=-\frac{\e\delta}{\tau}m.
    \label{evo-mean}
\end{equation}
As expected, the presence of a percentage cut $\delta>0$ in the jackpot game leads to an exponential decay to zero of the mean number of tickets at a rate proportional to $\frac{\e\delta}{\tau}$.
 
As far as higher order moments of the distribution function $f$ are concerned, analytic results may be obtained at the cost of more complicated computations, due to the non-linearity of the Boltzmann-type equation~\eqref{kine1}. This unpleasant fact is evident by computing, e.g. the second order moment, i.e. the energy of the system, which amounts to choosing $\varphi(x)=x^2$ in~\eqref{kine1}.  In this case, we have:
\begin{equation}
    \ave*{\sum_{k=1}^{N}{(x_k')}^2}=\left({(1-\e)}^2+2\e(1-\e)(1-\delta)\right)\sum_{k=1}^{N}x_k^2
        +\e^2{(1-\delta)}^2{\left(\sum_{k=1}^{N}x_k\right)}^2.
    \label{squ}
\end{equation}
Notice that the term $\bigl(\sum_{k=1}^{N}x_k\bigr)^2$, once integrated against the $N$-product of the distribution functions, produces a dependence on both the second moment and the square of the first moment, whose decay law has been established in~\eqref{evo-mean}.

It is now clear that, while giving a precise picture of the evolution of the jackpot game, the highly non-linear Boltzmann-type equation~\eqref{kine1} may essentially be treated only numerically.

\subsection{A linearised model}
\label{linea1}
A considerable simplification occurs in presence of a large number $N$ of participants in the game. In this situation, at any time $t>0$ we have
\begin{equation}
    \sum_{k=1}^{N}x_k=N\cdot\frac{1}{N}\sum_{k=1}^{N}x_k\approx Nm(t).
    \label{eq:approx.mean}
\end{equation}
In practice, if $N$ is large enough we may approximate the empirical mean number of tickets $\frac{1}{N}\sum_{k=1}^{N}x_k$ of the gamblers participating in a round of the game with the theoretical mean number of tickets $m$ owned by the entire population of potential gamblers. Hence, still considering for the moment the case $\beta=0$, the interaction~\eqref{micro} may be restated as
\begin{equation}
    x_k'=(1-\e)x_k+\e N(1-\delta)m(t)I(\tilde{A}-k), \qquad k=1,\,2,\,\dots,\,N,
    \label{micro1}
\end{equation}
where $\tilde{A}\in\{1,\,\dots,\,N\}$ is the discrete random variable with (approximate) law
$$ \P(\tilde{A}=k)\approx\frac{x_k}{Nm(t)}, \qquad k=1,\,\dots,\,N. $$

\begin{remark}
Owing to the approximation~\eqref{eq:approx.mean}, the usual properties $\P(\tilde{A}=k)\leq 1$ and $\sum_{k=1}^{N}\P(\tilde{A}=k)=1$ may be fulfilled, in general, only in a mild sense, which however becomes tighter and tighter as $N$ grows. We refrain from investigating precisely the proper order of magnitude of $N$, because, as we will see in a moment, we will be mostly interested in the asymptotic regime $N\to \infty$.
\end{remark}

Before proceeding further, we observe that in the recent paper~\cite{Bob11} the linearisation resulting from considering a large number of gamblers has been proposed in an economic context. The same type of approximation has also been used in~\cite{Demi} to linearise a Boltzmann-type equation describing the exchange of goods according to micro-economy principles.

The main consequence of the new interaction rule~\eqref{micro1} is that the each post-interaction number of tickets $x_k'$ depends linearly only on the pre-interaction number $x_k$ and on the (theoretical) mean number of tickets $m(t)$. Plugging~\eqref{micro1} into~\eqref{kine1} leads then to a linear Boltzmann-type equation. In particular, the time evolution of the observable quantities $\varphi=\varphi(x)$ is now given by
\begin{equation}
    \frac{d}{dt}\int_{\R_+}\varphi(x)f(x,\,t)\,dx=\frac{1}{\tau}\int_{\R_+}\ave{\varphi(x')-\varphi(x)}f(x,\,t)\,dx,
    \label{kine-lin}
\end{equation}
where
\begin{equation}
    x'=(1-\e)x+\e N(1-\delta)m(t)I(\bar{A}-1)
    \label{int4}
\end{equation}
and the random variable $\bar{A}\in\{0,\,1\}$ is such that
\begin{equation}
    \P(\bar{A}=1)=\frac{x}{Nm(t)}.
    \label{Ak11}
\end{equation}
In practice, since it is no longer necessary to label the single gamblers participating in a round of the jackpot game, we use $\bar{A}$ simply to decide whether the randomly chosen gambler $x$ wins ($\bar{A}=1$) or not ($\bar{A}=0$) in that round.

Equation~\eqref{kine-lin} allows for a simplified and explicit computation of the statistical moments of the distribution function $f$. In particular, it gives the right evolution of the first moment like in~\eqref{evo-mean}. We remark, however, that the simplified interaction rules~\eqref{int4}-\eqref{Ak11} have two main weak points. First, since the mean value $m(t)$ follows the decay given by~\eqref{evo-mean}, thus it is in particular non-constant in time, the interaction~\eqref{int4} features an explicit dependence on time. Second, if $\e$ is fixed independently of $N$, the number $\e N$ of tickets played in a single game tends to blow as $N$ increases. At that point, the kinetic model does not represent the target jackpot game any more. Therefore, while maintaining the fundamental linear characteristics, which make the model amenable to analytical investigations, it is essential to combine the large number of gamblers in each round with a simultaneously small value of $\e$. Indeed, it is realistic to assume that the product $\e N$, which characterises the percent number of tickets played in each game, remains finite for every $N\gg 1$ and $\e\ll 1$. We express this assumption by letting $\e\sim\kappa N^{-1}$, where $\kappa>0$ is a constant, so that
\begin{equation}
    \lim_{N\to\infty}\e N=\kappa.
    \label{bound1}
\end{equation}

\begin{remark}
Notice that the rate of decay of the mean value $m$ in the linear model~\eqref{kine-lin}, which, as already observed, equals the one of the non-linear model given by~\eqref{evo-mean}, is bounded away from zero for any value of $\e$ if and only if $\tau\sim\e$. Therefore, in order to maintain the correct decay of the mean value for any value of $\e$, $N$ in the linearised model, we will assume, without loss of generality, $\tau=\e$.
\label{rem:tau}
\end{remark}

We are now ready to re-include in the dynamics also the refilling of money operated by the gamblers drawing on their personal reserves of wealth. Assuming a very large number $N\gg 1$ of gamblers together with~\eqref{bound1} and taking also Remark~\ref{rem:tau} into account, the jackpot game with refilling is well described by the linear kinetic equation
\begin{equation}
    \frac{d}{dt}\int_{\R_+}\varphi(x)f(x,\,t)\,dx=
        \frac{1}{\e}\int_{\R_+^2}\ave{\varphi(x')-\varphi(x)}f(x,\,t)\Phi(y)\,dx\,dy,
    \label{kine2}
\end{equation}
where
\begin{equation}
    x'=(1-\e)x+\e\beta Y+\kappa(1-\delta)m(t)I(\bar{A}-1),
    \label{eq:micro.lin-refill}
\end{equation}
with $\bar{A}\in\{0,\,1\}$ and, recalling~\eqref{Ak11},
$$ \P(\bar{A}=1)=\e\frac{x}{\kappa m(t)}. $$
In~\eqref{kine2}, we denoted by $\Phi:\R_+\to\R_+$ the probability density function of the random variable $Y$ describing the refilling or money operated by the gamblers. Motivated by the discussion contained in the next Section~\ref{Behavior}, we assume that $\Phi$ is a lognormal probability density function. This agrees with the behaviour of the gamblers observed in~\cite{Wang} and ensures that the moments of $Y$ are all finite. In particular:
\begin{equation}
    M:=\int_{\R_+}y\Phi(y)\,dy<+\infty.
    \label{eq:M}
\end{equation}

Taking $\varphi(x)=x$ in~\eqref{kine2}, we obtain that the mean number of tickets owned by the gamblers obeys now the equation
$$ \frac{dm}{dt}=-\delta m+\beta M, $$
whence
\begin{equation}
    m(t)=m_0e^{-\delta t}+\frac{\beta M}\delta\left(1-e^{-\delta t}\right)
    \label{expl}
\end{equation}
with $m_0:=m(0)\geq 0$. Remarkably, $m$ does not depend on $\e$. Moreover, in presence of refilling, $m$ is uniformly bounded in time from above and from below:
$$ \min\left\{m_0,\,\frac{\beta M}\delta\right\}\leq m(t)\leq\max\left\{m_0,\,\frac{\beta M}\delta\right\}. $$
Note that, for $\beta,\,M>0$, the mean number of tickets $m$ no longer decays to zero but tends asymptotically to the value $\frac{\beta M}{\delta}$.

Choosing now $\varphi(x)=e^{-i\xi x}$, where $\xi\in\R$ and $i$ is the imaginary unit, we obtain the Fourier-transformed version of the kinetic equation~\eqref{kine2}:
\begin{equation}
    \partial_t\ff(\xi,\,t)=\frac{1}{\e}\int_{\R_+^2}\ave*{e^{-i\xi x'}-e^{-i\xi x}}f(x,\,t)\Phi(y)\,dx\,dy,
    \label{kine-f}
\end{equation}
where, as usual, $\ff$ denotes the Fourier transform of the distribution function $f$:
$$ \ff(\xi,\,t):=\int_{\R_+}f(x,\,t)e^{-i\xi x}\,dx. $$
Taking~\eqref{Ak11} into account, the right-hand side of~\eqref{kine-f} can be written as the sum of two contributions:
\begin{align*}
    A_\e(\xi,\,t) &= \frac{1}{\e}\int_{\R_+}\left(e^{-i\e\beta\xi y}-1\right)
        \left[\int_{\R_+}e^{-i\xi[(1-\e)x +\kappa(1-\delta)m(t)]}\frac{\e x}{\kappa m(t)}f(x,\,t)\,dx\right. \\
    &\phantom{=} +\left.\int_{\R_+}e^{-i(1-\e)\xi x}\left(1-\frac{\e x}{\kappa m(t)}\right)f(x,\,t)\,dx\right]\Phi(y)\,dy,
\end{align*}
and
\begin{align*}
    B_\e(\xi,\,t) &= \frac{1}{\e}\int_{\R_+}\left(e^{-i\xi[(1-\e)x+\kappa(1-\delta)m(t)]}-e^{-i\xi x}\right)\frac{\e x}{\kappa m(t)}f(x,\,t)\,dx \\
    &\phantom{=} +\frac{1}{\e}\int_{\R_+}\left(e^{-i(1-\e)\xi x}-e^{-i\xi x}\right)\left(1-\frac{\e x}{\kappa m(t)}\right)f(x,\,t)\,dx.
\end{align*}
In the limit $\e\to 0^+$, viz. $N\to\infty$, we obtain
\begin{align*}
    \lim_{\e\to 0^+}A_\e(\xi,\,t) &= -i\beta M\xi\ff(\xi,\,t) \\
    \lim_{\e\to 0^+}B_\e(\xi,\,t) &= \left[\frac{i}{\kappa m(t)}\left(e^{-i\kappa m(t)(1-\delta)\xi}-1\right)
        -\xi\right]\partial_\xi\ff(\xi,\,t),
\end{align*}
which shows that, for $N\gg 1$ and in the regime~\eqref{bound1}, the non-linear kinetic model~\eqref{kine1} with the scaling $\tau=\e$ (cf. Remark~\ref{rem:tau}) is well approximated by the Fourier-transformed linear equation
\begin{equation}
    \partial_t\ff=\left[\frac{i}{\kappa m(t)}\left(e^{-i\kappa m(t)(1-\delta)\xi}-1\right)-\xi\right]\partial_\xi\ff
        -i\beta M\xi\ff.
    \label{four}
\end{equation}
This equation may be used to compute recursively the time evolution of the statistical moments of $f$, upon recalling the relationship
\begin{equation}
    m_n(t):=\int_{\R_+}x^nf(x,\,t)\,dx=i^n\partial_\xi^n\ff(0,\,t), \quad n\in\N,
    \label{eq:Fourier.moments}
\end{equation}
and to check their possible blow up indicating the formation of fat tails in $f$.

\subsection{Explicit steady states and boundedness of moments}
\label{sect:gamma}
To gain further information on~\eqref{four} in the physical variable $x$, let us consider at first the case in which the constant $\kappa$ is small, say $\kappa\ll 1$. Expanding the exponential function appearing in~\eqref{four} in Taylor series up to the second order, we obtain
\begin{equation} 
    \left[\frac{i}{\kappa m(t)}\left(e^{-i\kappa m(t)(1-\delta)\xi}-1\right)-\xi\right]\partial_\xi\ff\approx
        \left[-\delta\xi-\frac{i\kappa m(t)}{2}(1-\delta)^2\xi^2\right]\partial_\xi\ff.
    \label{app1}
\end{equation}
Within this approximation, we can go back from~\eqref{four} to the physical variable $x$ by the inverse Fourier transform. In particular, we get
\begin{equation}
    \partial_tf(x,\,t)=\frac{\kappa(1-\delta)^2m(t)}{2}\partial_x^2(xf(x,\,t))+\partial_x\bigl((\delta x-\beta M)f(x,\,t)\bigr),
    \label{FPmod}
\end{equation}
which is a \textit{Fokker-Planck-type equation} with variable diffusion coefficient. Notice that the mean value of the solution to~\eqref{FPmod} coincides with~\eqref{expl}. In particular, if $m_0=\frac{\beta M}\delta$ then the mean value remains constant in time:
$$ m(t)\equiv\frac{\beta M}\delta \quad \forall\,t>0. $$
In this simple case,~\eqref{FPmod} has a stationary solution, say $f_\infty=f_\infty(x)$, which is easily found by solving the differential equation
$$ \frac{\kappa(1-\delta)^2}{2}\cdot\frac{\beta M}\delta\partial_x(xf_\infty)+(\delta x-\beta M)f_\infty=0 $$
and which turns out to be a \textit{gamma probability density function}:
\begin{equation}
    f_\infty(x)=\frac{{\left(\frac{2\delta^2}{\kappa(1-\delta)^2\beta M}\right)}^\frac{2\delta}{\kappa(1-\delta)^2}}{\Gamma\!\left(\frac{2\delta}{\kappa(1-\delta)^2}\right)}
        x^{\frac{2\delta}{\kappa(1-\delta)^2}-1}e^{-\frac{2\delta^2}{\kappa(1-\delta)^2\beta M}x}.
    \label{eq:gamma}
\end{equation}
Since  $f_\infty$ has moments bounded of any order, we conclude that \textit{no fat tail is produced in this case}.

In the general case, i.e. without invoking the approximation~\eqref{app1}, we may check that the same qualitative asymptotic trend emerges by resorting to the following argument. Let us define
$$ D(\xi,\,t):=\frac{i}{\kappa m(t)}\left(e^{-i\kappa m(t)(1-\delta)\xi}-1\right)-(1-\delta)\xi, $$
so that~\eqref{four} may be rewritten as
\begin{equation}
    \partial_t\ff=D(\xi,\,t)\partial_\xi\ff-\delta\xi\partial_\xi\ff-i\beta M\xi\ff.
    \label{four1}
\end{equation}
The function $D(\xi,\,t)$ satisfies
$$ D(0,\,t)=\partial_\xi D(0,\,t)=0, $$
while, for $n\geq 2$,
$$ \partial_\xi^nD(0,\,t)= {(i\kappa m(t))}^{n-1}{(1-\delta)}^n, $$
and further, owing to the Leibniz rule,
\begin{equation}
    \left.\partial_\xi^n\left(D(\xi,\,t)\partial_\xi\ff(\xi,\,t)\right)\right\vert_{\xi=0}
        =\sum_{k=2}^{n}\binom{n}{k}\partial_\xi^kD(0,\,t)\partial_\xi^{n-k+1}\ff(0,\,t).
    \label{nn}
\end{equation}
Notice that the highest order derivative of $\ff$ appearing on the right-hand side of~\eqref{nn} is of order $n-1$. Therefore, taking the $n$th $\xi$-derivative of~\eqref{four1} and computing in $\xi=0$, while recalling~\eqref{eq:Fourier.moments}, yields, for $n\geq 2$,
\begin{equation}
    \frac{dm_n}{dt}=-n\delta m_n+\mathcal{E}(m_1,\,\dots,\,m_{n-1}),
    \label{mom-n}
\end{equation}
where $\mathcal{E}$ is a term containing only moments of order at most equal to $n-1$. The exact expression of $\mathcal{E}$ may be obtained from~\eqref{eq:Fourier.moments}-\eqref{nn} but, in any case,~\eqref{mom-n} shows recursively that the statistical moments of $f$ of any order are uniformly bounded in time if they are bounded at the initial time. Therefore, we conclude that \textit{fat tails do not form} also in the general case described by~\eqref{four}.

\begin{remark}
The uniform boundedness of all moments of $f$ has been actually proved only for the linearised kinetic model~\eqref{kine-lin}-\eqref{int4} in the limit regime $\epsilon\to 0^+$, viz. $N\to\infty$. Nevertheless, the result so obtained suggests that also the ``real'' kinetic model, described by the highly non-linear Boltzmann-type equation~\eqref{kine1}, may behave in the same way. This is in contrast with the conclusions drawn in~\cite{Wang}, where, resorting to some simplified models, the authors justify the formation of power law tails in the distribution of the gambler winnings.

It is noticeable that equation~\eqref{four}, obtained in the limit of a very large number of gamblers participating in a round of the jackpot game, maintains all the essential features of the game. In particular, it preserves the fact that there is only one winner in each round. This imposes a strong correlation between the winnings of the gamblers, which clearly remains also in the limit. These characteristics are very close to those of the situation described in~\cite{BaTo}, where explicit steady states for a model of a pure gambling between two players are found. Specifically, if in each round there is exactly one winner and one loser then it is proved that the steady state possesses all moments bounded. Conversely, if both gamblers may simultaneously win or lose in a round then power law tails appear at equilibrium.
\label{no-fat}
\end{remark}

\subsection{Are power law tails correct?}
\label{fat}
As briefly outlined in Remark \ref{no-fat}, the solution to the linearised kinetic model of the jackpot game does not possess fat tails. In order to investigate the possible reasons behind the fat tails apparently observed in~\cite{Wang}, in the following we introduce an alternative linear kinetic model of the jackpot game, still derived from the microscopic interaction~\eqref{micro}, whose equilibrium density exhibits indeed power-law-type fat tails. This new model may be obtained by resorting to a different linearisation of~\eqref{kine1}. Nevertheless, as observed via numerical experiments in the next Section~\ref{Numerics}, such a linearised equation, while apparently very close to the original non-linear model, produces a quite different large-time trend compared to the one described by~\eqref{four}.

Let us fix $\beta=0$ in~\eqref{micro} and let us assume, without loss of generality, that the extracted winner is the gambler $k=1$. Then:
\begin{align*}
    x_1' &= (1-\e)x_1+\e(1-\delta)\sum_{j=1}^{N}x_j \\
    x_k' &= (1-\e)x_k, & k=2,\,3,\,\dots,\,N,
\end{align*}
which implies (cf. also~\eqref{squ}):
$$ \sum_{k=1}^{N}{(x_k')}^2={(1-\e)}^2\sum_{k=1}^{N}x_k^2
    +2\e(1-\e)(1-\delta)x_1\sum_{k=1}^{N}x_k
    +\e^2{(1-\delta)}^2{\left(\sum_{k=1}^{N}x_k\right)}^2. $$
Taking into account the expression \eqref{mean} of the mean value, we obtain
\begin{align*}
    \frac{N\sum\limits_{k=1}^{N}{(x_k')}^2}{{\left(\sum\limits_{k=1}^{N}x_k'\right)}^2} &=
        \frac{{(1-\e)}^2}{{(1-\e\delta)}^2}\frac{N\sum\limits_{k=1}^{N}x_k^2}{\left(\sum\limits_{k=1}^{N}x_k\right)^2}
            +N\e^2{(1-\delta)}^2+2N\e(1-\e)(1-\delta)\frac{x_1}{\sum\limits_{k=1}^{N}x_k} \\
    &\approx \frac{N\sum\limits_{k=1}^{N}x_k^2}{{\left(\sum\limits_{k=1}^{N}x_k\right)}^2}
 \end{align*}
for $N\gg 1$ large and, consequently, $\epsilon\ll 1$ small. Indeed,
$$ \frac{x_1}{\sum\limits_{k=1}^{N}x_k}=\frac{\frac{1}{N}x_1}{\frac{1}{N}\sum\limits_{k=1}^{N}x_k}
    \approx\frac{x_1}{Nm(t)}\xrightarrow{N\to\infty} 0. $$
In other words, for a large number $N$ of gamblers and a correspondingly small percentage $\e$ of tickets played in a single game, the relationship~\eqref{bound1} implies that the quantity 
\begin{equation}
    \chi:=\frac{N\sum\limits_{k=1}^{N}x_k^2}{{\left(\sum\limits_{k=1}^{N}x_k\right)}^2}
    \label{eq:chi}
\end{equation}
may be regarded approximately as a \emph{collision invariant} of the interaction~\eqref{micro}. Since
$$ {\left(\sum_{k=1}^{N}x_k\right)}^2\leq N\sum_{k=1}^{N}x_k^2, $$
it follows that $\chi\geq 1$. Note that this result does not depend on the choice of the winner in each round of the jackpot game.

Using~\eqref{bound1} and~\eqref{eq:chi} in~\eqref{squ}, in this asymptotic approximation we obtain:
\begin{align}
    \begin{aligned}[b]
        \ave*{\sum_{k=1}^{N}{(x_k')}^2} &= \left({(1-\e)}^2+2\e(1-\e)(1-\delta)\right)\sum_{k=1}^{N}x_k^2
            +\e(1-\delta)^2\frac{\kappa}{\chi}\sum_{k=1}^{N}x_k^2 \\
        &= \left[{(1-\e\delta)}^2+\e{(1-\delta)}^2\left(\frac{\kappa}{\chi}-\e\right)\right]\sum_{k=1}^{N}x_k^2,
    \end{aligned}
    \label{squ-c}
\end{align}
whence, choosing $\varphi(x)=x^2$ in~\eqref{kine1},
\begin{equation}
    \frac{dm_2}{dt}=\left[\e\left({(1-\delta)}^2\frac{\kappa}{\chi}-2\right)-\e^2(1-2\delta)\right]m_2.
    \label{evo-square}
\end{equation}
This equation shows that the ratio $\kappa/\chi$ is of paramount importance to classify the large-time trend of the energy of the distribution $f$, hence also of $f$ itself. Indeed, the sign of the coefficient
$$ c(\kappa,\,\chi,\,\e):=\e\left({(1-\delta)}^2\frac{\kappa}{\chi}-2\right)-\e^2(1-2\delta) $$
determines if $f$ converges asymptotically in time to a Dirac delta centred in $x=0$ (when $c(\kappa,\,\chi,\,\e)<0$) or if it spreads on the whole positive real line (when $c(\kappa,\,\chi,\,\e)>0$).

This discussion suggests a consistent way to eliminate the time dependence in the interaction~\eqref{int4}, while preserving the main \textit{macroscopic} properties of the jackpot game, such as the right time evolutions of the mean, cf.~\eqref{evo-mean}, and of the energy, cf.~\eqref{evo-square}. Specifically, we proceed as follows. For all observable quantities $\varphi=\varphi(x)$, we consider the linear kinetic model~\eqref{kine-lin} with the following linear interaction rule:
\begin{equation}
    x'=(1-\e\delta)x+\sqrt{\e}x\eta_\e,
    \label{new}
\end{equation}
where $\e>0$. In~\eqref{new}, $\eta_\e$ is a discrete random variable taking only the two values $-\sqrt{\e}(1-\delta)$, $M_\e/\sqrt{\e}$ with probabilities
$$ \P\!\left(\eta_\e=-\sqrt{\e}(1-\delta)\right)=1-p_\e, \qquad 
        \P\!\left(\eta_\e=\frac{M_\e}{\sqrt{\e}}\right)=p_\e, $$
where $p_\e\in [0,\,1]$ and $M_\e>0$ are two constants to be properly fixed.

We interpret the rule~\eqref{new}, together with the prescribed values of $\eta_\e$, as follows: a gambler, who enters the game with a number of tickets (viz. an amount of money) equal to $\e x$, may either win a jackpot equal to $(M_\e-\e\delta)x$ with probability $p_\e$ or lose the amount $\e x$ put into the game with probability $1-p_\e$.

In particular, we determine $p_\e$ by imposing $\ave{\eta_\e}=0$, which guarantees that~\eqref{new} reproduces the correct evolution of the mean provided by~\eqref{int4} (indeed, in such a case we have $\ave{x'}=(1-\e\delta)x$). We find then
$$ p_\e=\frac{\e(1-\delta)}{M_\e+\e(1-\delta)}. $$
Using this, we discover $\ave{\eta_\e^2}=M_\e(1-\delta)$, whence
\begin{equation}
    \ave{{(x')}^2}=\left({(1-\e\delta)}^2+\e(1-\delta)M_\e\right)x^2.
    \label{m22}
\end{equation}
A comparison between formulas~\eqref{squ-c} and~\eqref{m22} allows us to conclude that the choice
$$ M_\e=(1-\delta)\left(\frac\kappa\chi-\e\right) $$
further implies a time evolution of the energy identical to~\eqref{evo-square}. Notice that the positivity of $M_\e$ is guaranteed by choosing $\e\ll 1$ small enough.

After deriving the linearised model for $\beta=0$, we may re-include the refilling of tickets/money in the interaction rule:
\begin{equation}
    x'=(1-\e\delta)x+\e\beta Y+\sqrt{\e}x\eta_\e,
    \label{new1}
\end{equation}
where, as stated in Section~\ref{linea1}, the random variable $Y\in\R_+$ is described by a prescribed lognormal probability density function $\Phi:\R_+\to\R_+$.

Within this approximation of the dynamics, the evolution of the distribution function $g=g(x,\,t)$ of the tickets (viz. the money) played and won by a large number of gamblers participating in the jackpot game is then described by the linear kinetic equation (cf. also~\eqref{kine2}):
\begin{equation}
    \frac{d}{dt}\int_{\R_+}\varphi(x)g(x,\,t)\,dx=
        \frac{1}{\tau}\int_{\R^2_+}\ave{\varphi(x')-\varphi(x)}g(x,\,t)\Phi(y)\,dx\,dy
    \label{kine-full}
\end{equation}
with $x'$ given by~\eqref{new1}.

\subsubsection{Fokker-Planck description of the jackpot game}
\label{sect:FP.jackpot}
The linear kinetic equation~\eqref{kine-full} describes the evolution of the distribution function due to interactions of type~\eqref{new1}. As discussed in Section~\ref{linea1}, for large values of the number $N$ of gamblers participating in a round, and therefore, in view of~\eqref{bound1}, a small value of $\e$, the interaction~\eqref{new1} produces a small variation in the number of tickets owned by a gambler. We say then that, in such a regime, the interaction~\eqref{new1} is \textit{quasi-invariant} or \textit{grazing}. Consequently, a finite (i.e., non-infinitesimal) evolution of {the distribution function $g$} may be observed only if each gambler participates in a huge number of interactions~\eqref{new1} during a fixed period of time. This is achieved by means of the scaling $\tau\sim\e$ like in Section~\ref{linea1}, cf. Remark~\ref{rem:tau}.
 
In this scaling, the kinetic model~\eqref{kine-full} is shown to approach its continuous counterpart given by a Fokker-Planck-type equation~\cite{FPTT,pareschi2013BOOK,Villani}. In the present case,~\eqref{kine-full} is well approximated by the following weak form of a new linear Fokker-Planck equation with variable coefficients:
\begin{equation}
    \frac{d}{dt}\int_{\R_+}\varphi(x)g(x,\,t)\,dx=
        \int_{\R_+}\left(-\varphi'(x)(\delta x-\beta M)+\frac{\tilde{\sigma}}{2}\varphi''(x)x^2\right)g(x,\,t)\,dx,
      \label{m-15}
\end{equation}
where $M$ is the mean refilling of tickets, cf.~\eqref{eq:M}, and where we have defined
$$ \tilde{\sigma}:=\lim_{\e\to 0^+}M_\e=(1-\delta)\frac{\kappa}{\chi}. $$
Then, provided the boundary terms produced by the integration by parts vanish,~\eqref{m-15} may be recast in strong form as
\begin{equation}
    \partial_tg(x,\,t)=\frac{\tilde{\sigma}}{2}\partial^2_x(x^2g(x,\,t))+\partial_x((\delta x-\beta M)g(x,\,t)).
    \label{FP3}
 \end{equation}
This equation describes the evolution of the distribution function $g$ of the number of tickets $x\in\R_+$ played by the gamblers at time $t>0$ in the limit of the \textit{grazing} interactions. The advantage of this equation over~\eqref{kine-full} is that its unique steady state $g_\infty$ with unitary mass may be explicitly computed:
\begin{equation}
    g_\infty(x)=\frac{\left(\frac{2\beta M}{\tilde{\sigma}}\right)^{1+\frac{2\delta}{\tilde{\sigma}}}}{\Gamma\!\left(1+\frac{2\delta}{\tilde{\sigma}}\right)}\cdot
            \frac{e^{-\frac{2\beta M}{\tilde{\sigma}x}}}{x^{2+\frac{2\delta}{\tilde{\sigma}}}}.
    \label{eq:invgamma}
\end{equation}
We observe that this is an \textit{inverse gamma} probability density function with parameters linked to the details of the microscopic interaction~\eqref{new1}.

\begin{remark}
A comparison between~\eqref{FPmod} and the Fokker-Planck equation~\eqref{FP3} shows that, while the drift term is the same, the coefficient of the diffusion term is proportional to $x$ in~\eqref{FPmod} and to $x^2$ in~\eqref{FP3}. This difference determines, in the latter case, the formation of \textit{fat tails}, which is consistent with the claim made in~\cite{Wang}. Nevertheless, as briefly explained before, the approach based on the interaction~\eqref{new1} leading to~\eqref{FP3} in the quasi-invariant regime does \textit{not} actually describe exactly the jackpot game. Indeed, it admits that all gamblers may win simultaneously, although with a very small probability.
\end{remark}

\subsubsection{The case~\texorpdfstring{$\boldsymbol{\beta=0}$}{}}
Further explicit computations on the Fokker-Planck equation~\eqref{FP3} may be done in the case $\beta=0$, which corresponds to the situation in which gamblers enter the game with a certain number of tickets, viz. amount of money, and use only those tickets, viz. money, to play. Then, the distribution function $g=g(x,\,t)$ solves the equation
\begin{equation}
    \partial_tg(x,\,t)=\frac{\tilde{\sigma}}{2}\partial^2_x(x^2g(x,\,t))+\delta\partial_x(xg(x,\,t)).
    \label{FPnull}
\end{equation}
Setting
$$ \tilde{g}(x,\,t):=e^{-\delta t}g(e^{-\delta t}x,\,t), $$
which is easily checked to be in turn a distribution function with unitary mass at each time $t>0$, we see that $\tilde{g}$ solves the diffusion equation
\begin{equation}
    \partial_t\tilde{g}(x,\,t)=\frac{\tilde{\sigma}}{2}\partial^2_x(x^2\tilde{g}(x,\,t))
    \label{geo}
\end{equation}
with the same initial datum as that prescribed to~\eqref{FPnull}, because $\tilde{g}(x,\,0)=g(x,\,0)$.

The unique solution to~\eqref{geo} corresponding to an initial datum $g_0(x)$ is given by the expression:
\begin{equation}
    \tilde{g}(x,\,t)=\int_{\R_+}\frac{1}{z}g_0\!\left(\frac{x}{z}\right)L_t(z)\,dz,
    \label{sol-geo}
\end{equation}
where
$$ L_t(x):=\frac{1}{\sqrt{2\pi\tilde{\sigma}t}x}\exp\!{\left(-\frac{{(\log{x}+\frac{\tilde{\sigma}}{2}t)}^2}{2\tilde{\sigma}t}\right)} $$
is a lognormal probability density. Indeed,~\eqref{geo} possesses a unique source-type solution given by a lognormal density with unit mean, which at time $t=0$ coincides with a Dirac delta centred in $x=1$, cf.~\cite{Tos16}.
 
Both the mass and the mean of~\eqref{sol-geo} are conserved in time, while initially bounded moments of order $n\geq 2$ grow exponentially at rate $n(n-1)$. Moreover,~\eqref{sol-geo} can be shown to converge in time to $L_t(x)$ in various norms, see~\cite{Tos16}.

Starting from~\eqref{sol-geo}, we easily obtain that the unique solution to the original Fokker-Planck equation~\eqref{FPnull} is given by
\begin{equation}
    g(x,\,t)=\int_{\R_+}\frac{1}{z}g_0\!\left(\frac{x}{z}\right)\tilde{L}_t(z)\,dz,
    \label{sol-or}
\end{equation}
where
\begin{equation}
    \tilde{L}_t(x)=\frac{1}{\sqrt{2\pi\tilde{\sigma}t}x}
        \exp\!{\left(-\frac{{\bigl(\log{x}+(\delta+\frac{\tilde{\sigma}}{2})t\bigr)}^2}{2\tilde{\sigma}t}\right)}.
    \label{ln2}
\end{equation}
Notice that, as expected, the mean value of the lognormal density~\eqref{ln2} decays exponentially in time:
$$ \int_{\R_+}x\tilde{L}_t(x)\,dx=e^{-\delta t}. $$
Consequently, if $X_t\sim g(x,\,t)$ is a stochastic process with probability density equal to the solution of~\eqref{FPnull}, the mean of $X_t$ decays exponentially to zero at the same rate and
$$ \ave{X_t}=\int_{\R_+}xg(x,\,t)\,dx=e^{-\delta t}\ave{X_0}. $$
Taking advantage of the representation formula~\eqref{sol-or}, we can easily compute also higher order moments of the solution. In particular, the variance of $X_t$ is equal to
$$ \ave{X_t^2}-\ave{X_t}^2=\ave{X_0^2}e^{(\tilde{\sigma}-2\delta)t}-\ave{X_0}^2e^{-2\delta t}. $$
From here, we see that the large time trend of the variance depends on the sign of the quantity $\tilde{\sigma}-2\delta$. If $\tilde{\sigma}<2\delta$, the variance converges exponentially to zero, thus all gamblers tend, in the long run, to lose all their tickets (viz. money). Conversely, if $\tilde{\sigma}>2\delta$, the variance blows up for large times. This situation is analogous to the \textit{winner takes it all} behaviour~\cite{pareschi2013BOOK}, where the asymptotic steady state is a Dirac delta centred in zero but at any finite time a small decreasing number of gamblers possesses a huge number of tickets, sufficient to sustain the growth of the variance.

\section{Agent behaviour on gambling}
\label{Behavior}
A non-secondary aspect of the online gambling is related to the behavioural trends of the gamblers. The data analysis in~\cite{Wang} focuses, in particular, on two characteristics of the gambling activity: first, the \textit{waiting time}, defined as the time, measured in seconds, between successive bets by the same gambler; second, the \textit{number of rounds} played by individual gamblers. The study of this second aspect may shed light on the reasons behind a high gambling frequency and therefore also on possible addiction problems caused by gambling.

The fitting of the number of rounds played by individual gamblers during the period covered by gambling logs allowed the authors of~\cite{Wang} to conclude that the number of rounds is well described by a \textit{lognormal distribution}. This result is in agreement with other studies, cf. e.g.~\cite{Lund2008} and references therein, where the mean gambling frequency is put in close relation with the alcohol consumption. Starting from the pioneering contribution~\cite{Led56}, it has long been acknowledged that there exists a positive correlation between the level of alcohol consumption in a population and the proportion of heavy drinkers in the society. This relationship is known under several names, such as the \textit{total consumption model} or \textit{the single distribution theory}. Previous research has also found that its validity is not limited to the alcohol consumption but extends to different human phenomena.

In some recent papers~\cite{Gua1,Gua2}, we introduced a kinetic description of a number of human behavioural phenomena, which recently has been applied also to the study of alcohol consumption~\cite{DT19}. The modelling assumptions in~\cite{DT19} allowed us to classify the alcohol consumption distribution as a generalised gamma probability density, which includes the lognormal distribution as a particular case. Recalling that, as discussed above, alcohol consumption shows a lot of similarities with the gambling activity and taking inspiration from~\cite{DT19,Gua2}, we may explain exhaustively two main phenomena linked to the gambler behaviour. On one hand, the distribution of the number of tickets which individual gamblers play (including the refilling) in a single round of the jackpot game. On the other hand, the distribution of the number of rounds played by individual gamblers in time. Concerning this second aspect, a fitting of empirical data is presented and analysed in~\cite{Wang}. Conversely, no mention is made therein about the first aspect. For this reason, in the following we will be mainly interested in the problem of the distribution of the number of tickets used by gamblers in each round, which provides the law of the $Y_k$'s appearing in~\eqref{n-gen},~\eqref{micro} and of $Y$ appearing in~\eqref{eq:micro.lin-refill}. From the discussion about it, it will be possible to draw conclusions also on the problem of the number of rounds played by individual gamblers, since both problems are actually subject to identical microscopic rules, cf. Remark~\ref{rem:rounds} below.

\subsection{Kinetic modelling and value functions}
The  evolution of the number density of tickets which the gamblers purchase to participate in successive rounds of the jackpot game may be still treated resorting to the principles of statistical mechanics. Specifically, one can think of the population of gamblers as a multi-agent system: each gambler undergoes a sequence of microscopic interactions, through which s/he updates the personal number of tickets. In order to keep the connection with the classical kinetic theory of rarefied gases, these interactions obey suitable and universal rules, which, in the absence of well-defined physical laws, are designed so as to take into account at best some of the psychological aspects related to gambling.

Due to the nature of the game, the players know that there is a high probability to lose and a small one to win. For this reason, they are usually prepared to participate in a sequence of rounds, hoping to win in at least one of them. The involvement in the game pushes the gamblers to participate in successive rounds by purchasing an increasing number of tickets, so as to increase the probability to win. On the other hand, the attempt to safeguard the personal wealth suggests them to fix an \textit{a priori} upper bound to the number of tickets purchased. These two aspects, clearly in conflict, are characteristic of a typical human behaviour, which has been recently modelled in similar situations~\cite{DT19,Gua1,Gua2}. There, the microscopic interactions have been built taking inspiration from the pioneering analysis by Kahneman and Twersky~\cite{KT} about decisional processes under risk.

In the present case, the aforementioned safeguarding tendency may be modelled by assuming that the gamblers have in mind an ideal number $\bar{w}>0$ of tickets to buy in each round and, simultaneously, a threshold $\bar{w}_L>\bar{w}$, which they had better not exceed in order to avoid a (highly probable) excessive loss of money. Hence, the natural tendency of the gamblers to increase their number of tickets $w>0$ bought for the forthcoming rounds has to be coupled with the limit value $\bar{w}_L$, which it would be wise not to exceed. Following~\cite{Gua1,Gua2}, we may realise a gambler update via the following rule:
\begin{equation}
    w'=w-\Psi\left(\frac{w}{\bar{w}_L}\right)w+w\eta.
    \label{coll}
\end{equation}
In~\eqref{coll}, $w$, $w'$ denote the numbers of tickets played in the last round and in the forthcoming one, respectively. The function $\Psi$ plays the role of the so-called \textit{value function} in the prospect theory by Kahneman and Twersky~\cite{KT}. Specifically, it determines the update of the number of tickets in a \textit{skewed} way, so as to reproduce the behavioural aspects discussed above. Analogously to~\cite{Gua1}, we let
\begin{equation}
    \Psi(s):=\mu\frac{s^\alpha-1}{s^\alpha+1}, \qquad  s\geq 0,
    \label{vf}
\end{equation}
where $\mu,\,\alpha\in (0,\,1)$ are suitable constants characterising the agent behaviour. In particular, $\mu$ denotes the maximum variation in the number of tickets allowed in a single interaction~\eqref{coll}, indeed
\begin{equation}
    \abs{\Psi(s)}\leq\mu \quad \forall\,s\geq 0.
    \label{bounds}
\end{equation}
Hence, a small value of $\mu$ describes gamblers who buy a regular number of tickets in each round.

\begin{figure}[!t]
\centering
\includegraphics[width=0.6\textwidth]{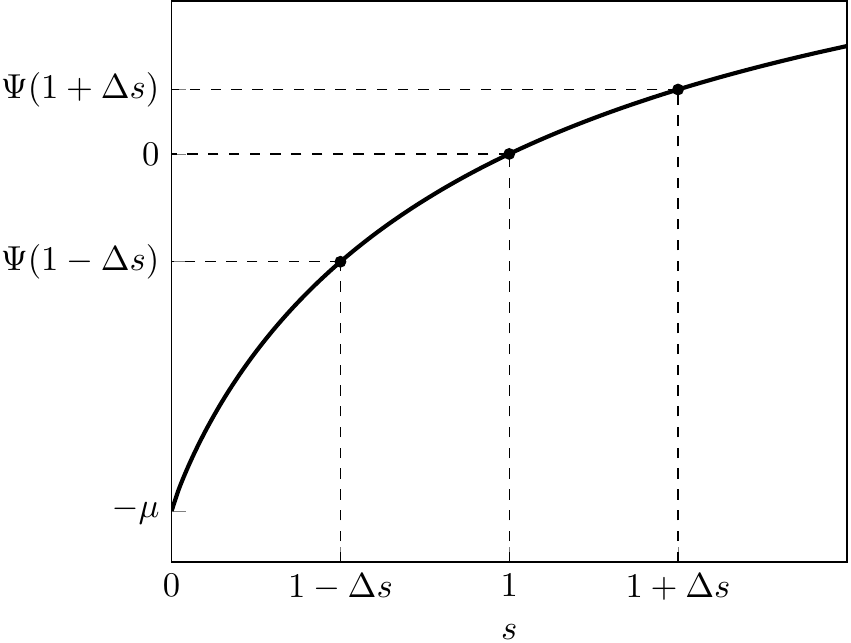}
\caption{The function $\Psi$ given in~\eqref{vf}.}
\label{fig:vf}
\end{figure}

The function $\Psi$ given in~\eqref{vf} maintains most of the physical properties required to the value function in the prospect theory~\cite{KT} and is particularly suited to the present situation. In the microscopic interaction~\eqref{coll}, the minus sign in front of $\Psi$ is related to the fact that the desire to increase the probability to win pushes a gambler to increase the number $w$ of purchased tickets when $w<\bar{w}_L$. At the same time, the tendency to safeguard the personal wealth induces the gambler to reduce the number of purchased tickets when $w>\bar{w}_L$. Moreover, the function $\Psi$ is such that
$$ -\Psi(1-\Delta{s})>\Psi(1+\Delta{s}), \qquad \forall\,\Delta{s}\in (0,\,1), $$
cf. Figure~\ref{fig:vf}. This inequality means that, if two gamblers are at the same distance from the limit value $\bar{w}_L$ from below and from above, respectively, the gambler starting from below will move closer to the optimal value $\bar{w}_L$ than the gambler starting from above. In other words, it is typically easier for a gambler to allow her/himself to buy more tickets, when the optimal threshold has not been exceeded, than to limit her/himself, when the optimal threshold has already been exceeded.

Finally, in order to take into account a certain amount of human unpredictability in buying tickets in a new round, it is reasonable to assume that the new number of tickets may be affected by random fluctuations, expressed by the term $w\eta$ in~\eqref{coll}. Specifically, $\eta$ is a centred random variable
$$ \ave{\eta}=0, \qquad \ave{\eta^2}=\lambda>0, $$
meaning that the random fluctuations are negligible on average. Moreover, to be consistent with the necessary non-negativity of $w'$, we assume that $\eta>-1+\mu$, i.e. that the support of $\eta$ is bounded from the left.

\begin{remark}
The behaviour modelled by~\eqref{coll}, which in principle concerns only the losers, may actually be applied also to the unique winner. Indeed, if the winner remains into the game, the pleasure to play will be dominant, so that it is reasonable to imagine that the future behaviour will not depend too much on the number of tickets gained in the last round.
\end{remark}

\begin{remark}
As discussed in~\cite{Gua1}, the function~\eqref{vf} may be modified to better match the phenomenon under consideration. For example, in order to differentiate the rates of growth and of decrease of $\Psi$ and to stress the difficulty of the gamblers to act against such a skewed trend, one may consider the following modified value function:
$$ \Psi(s)=\mu\frac{s^\alpha-1}{\nu s^\alpha+1}, \qquad  s\geq 0, $$
with $\nu>1$, so that the bounds~\eqref{bounds} modify to
$$ -\mu\leq\Psi(s)\leq\frac{\mu}{\nu}<\mu. $$
In this case, the possibility to go against the natural tendency is slowed down.
Also, as discussed in~\cite{DT19}, the shape of the value function~\eqref{vf} can be generalised so as to better take into account possible addiction effects. The general class of value functions considered there is given by
\begin{equation}
    \Psi(s)=\mu\frac{e^{(s^\delta-1)/\delta}-1}{e^{(s^\delta-1)/\delta}+1}, \qquad  s\geq 0,
    \label{dif-vf}
\end{equation}
where $\delta\in (0,\,1]$ is a constant. This choice leads to different skewed steady states, in the form of generalised gamma densities.
\end{remark}

\begin{remark}
The discussion set forth applies also to the modelling of the number of rounds played by individual gamblers in a fixed period of time, which has been considered in~\cite{Wang}. In particular, we may assume that the gamblers establish \textit{a priori} to play for a limited number of times, in order to spend only a certain total amount of money. But then, as it happens in the single game, it is more difficult to stop than to continue. This can be well described by the rule~\eqref{coll} and by the value function~\eqref{vf}, where now $w$ represents the number of rounds played in the time period.
\label{rem:rounds}
\end{remark}

Let now $h=h(w,\,t)$ be the distribution function of the number of tickets purchased by a gambler in a certain round of the jackpot game. As anticipated at the beginning of this section, its time evolution may be obtained by resorting to kinetic collision-like models~\cite{pareschi2013BOOK} based on~\eqref{coll}. In particular, since the interaction~\eqref{coll} depends only on the behaviour of a single gambler, $h$ obeys a linear Boltzmann-type equation of the form
\begin{equation}
    \frac{d}{dt}\int_{\R_+}\varphi(w)h(w,\,t)\,dw=
        \frac{1}{\tau}\int_{\R_+}\ave{\varphi(w')-\varphi(w)}h(w,\,t)\,dw,
    \label{kin-w}
\end{equation}
cf.~\eqref{kine-lin}, where the constant $\tau>0$ measures the interaction frequency and $\varphi$ is any observable quantity.

Since the elementary interaction~\eqref{coll} is non-linear with respect to $w$, the only conserved quantity in~\eqref{kin-w} is obtained from $\varphi(w)=1$:
$$ \frac{d}{dt}\int_{\R_+}h(w,\,t)\,dw=0, $$
which implies that the solution to~\eqref{kin-w} remains a probability density at all times $t>0$ if it is so at the initial time $t=0$. The evolution of higher order moments is difficult to compute explicitly. As a representative example, let us take $\varphi(w)=w$, which provides the evolution of the mean number of tickets purchased by the gamblers over time:
$$ m(t):=\int_{\R_+}wh(w,\,t)\, dw. $$
Since
$$ \ave{w'-w}=\mu\frac{w^\alpha-\bar{w}_L^\alpha}{w^\alpha+\bar{w}_L^\alpha}w, $$
we obtain
\begin{equation}
    \frac{dm}{dt}=\frac{\mu}{\tau}\int_{\R_+}\frac{w^\alpha-\bar{w}_L^\alpha}{w^\alpha+\bar{w}_L^\alpha}wh(w,\,t)\,dw.
    \label{evo-m}
\end{equation}
This equation is not explicitly solvable. However, in view of~\eqref{bounds}, $m$ remains bounded at any time $t>0$ provided it is so initially, with the explicit upper bound, cf.~\cite{Gua2},
$$ m(t)\leq m_0e^{\frac{\mu}{\tau}t}, $$
where $m_0:=m(0)$. From~\eqref{evo-m} it is however not possible to deduce whether the time variation of $m$ is or is not monotone.

Taking now $\varphi(w)=w^2$ in~\eqref{kin-w} and considering that
$$ \ave{{(w')}^2-w^2}=\left(\Psi^2\left(\frac{w}{\bar{w}_L}\right)-2\Psi\left(\frac{w}{\bar{w}_L}\right)
    +\lambda\right)w^2\leq (3\mu+\lambda)w^2 $$
because of~\eqref{bounds} together with $0<\mu<1$, we see that the boundedness of the energy at the initial time implies that of the energy at any subsequent time $t>0$, with the explicit upper bound
$$ m_2(t)\leq m_{2,0}e^{\frac{{ 3\mu+\lambda}}{\tau}t}, $$
where $m_{2,0}:=m_2(0)$.

\subsection{Fokker-Planck description and equilibria}
The linear kinetic equation~\eqref{kin-w} is valid for every choice of the parameters $\alpha$, $\mu$ and $\lambda$, which characterise the microscopic interaction~\eqref{coll}. In real situations, however, a single interaction, namely a participation in a new round of the jackpot game, does not induce a marked change in the value of $w$. This situation is close to that discussed in Section \ref{sect:FP.jackpot}, where we called these interactions \textit{grazing collisions}~\cite{pareschi2013BOOK,Villani}. 

Similarly to Section~\ref{sect:FP.jackpot}, we may easily take such a smallness into account by scaling the microscopic parameters in~\eqref{coll},~\eqref{kin-w} as
\begin{equation}
    \alpha\to\e\alpha, \qquad \lambda\to\e\lambda, \qquad \tau=\e,
    \label{scal}
\end{equation}
where $0<\e\ll 1$. A thorough discussion of these scaling assumptions may be found in~\cite{FPTT,Gua1}. In particular, here we mention that the rationale behind the coupled scaling of the parameters $\alpha$, $\lambda$ and of the frequency of the interactions $\tau$ is the following: since the scaled interactions are grazing, and consequently produce a very small change in $w$, a finite (i.e. non-infinitesimal) variation of the distribution function $g$ may be observed only if each gambler participates in a very large number of interactions within a fixed period of time.

As already observed in Section~\ref{sect:FP.jackpot}, when grazing interactions dominate, the kinetic model~\eqref{kin-w} is well approximated by a Fokker-Planck type equation~\cite{pareschi2013BOOK,Villani}. Exhaustive details on such an approximation in the kinetic theory of socio-economic systems may be found in~\cite{FPTT}. In short, the mathematical idea is the following: if $\varphi$ is sufficiently smooth and $w'\approx w$ because interactions are grazing, one may expand $\varphi(w')$ in Taylor series about $w$. Plugging such an expansion into~\eqref{kin-w} with the value function~\eqref{vf} and taking the scaling~\eqref{scal} into account one obtains:
$$ \frac{d}{dt}\int_{\R_+}\varphi(w)h(w,\,t)\,dw=
    \int_{\R_+}\left(-\frac{\alpha\mu}{2}\varphi'(w)w\log{\frac{w}{\bar{w}_L}}+
        \frac{\lambda}{2}\varphi''(w)w^2\right)h(w,\,t)\,dw+\frac{1}{\e}\mathcal{R}_\e(w,\,t), $$
where $\mathcal{R}_\e$ is a remainder such that $\frac{1}{\e}\mathcal{R}_\e\to 0$ as $\e\to 0^+$, cf.~\cite{FPTT}. Therefore, under the scaling~\eqref{scal}, the kinetic equation~\eqref{kin-w} is well approximated by the equation
$$ \frac{d}{dt}\int_{\R_+}\varphi(w)h(w,\,t)\,dw=
    \int_{\R_+}\left(-\frac{\alpha\mu}{2}\varphi'(w)w\log{\frac{w}{\bar{w}_L}}
        +\frac{\lambda}{2}\varphi''(w)w^2\right)h(w,\,t)\,dw. $$
This equation may be recognised as the weak form of the following Fokker-Planck equation with variable coefficients:
\begin{equation}
    \partial_t h(w,\,t)=\frac{\lambda}{2}\partial^2_w(w^2h(w,\,t))+\frac{\alpha\mu}{2}\partial_w\left(w\log{\frac{w}{\bar{w}_L}}h(w,\,t)\right),
    \label{FP2}
\end{equation}
upon assuming that the boundary terms produced by the integration by parts vanish. Like in Section~\ref{sect:FP.jackpot}, the Fokker-Planck description~\eqref{FP2} is advantageous over the original Boltzmann-type equation~\eqref{kin-w} because it allows for an explicit computation of the steady state distribution function, say $h_\infty=h_\infty(w)$. The latter solves the following first order ordinary differential equation:
$$ \frac{\lambda}{2}\frac{d}{dw}(w^2h_\infty(w))+\frac{\alpha\mu}{2}w\log{\frac{w}{\bar{w}_L}}h_\infty(w)=0, $$
whose unique solution with unitary mass is
\begin{equation}
    h_\infty(w)=\frac{1}{\sqrt{2\pi\sigma}w}\exp\left(-\frac{{(\log{w}-\theta)}^2}{2\sigma}\right),
    \label{eq:hinf_log}
\end{equation}
where 
$$ \sigma:=\frac{\lambda}{\alpha\mu}, \qquad \theta:=\log{\bar{w}_L}-\sigma. $$
Therefore, in very good agreement with the observations made in~\cite{Wang}, the equilibrium distribution function predicted by the microscopic rule~\eqref{coll} with the value function~\eqref{vf} in the grazing interaction regime is a \textit{lognormal probability density}, whose mean and variance are easily computed from the known formulas for lognormal distributions:
$$ m_\infty:=\bar{w}_Le^{-\frac{\sigma}{2}}, \qquad
    \operatorname{Var}(g_\infty):=\bar{w}_L^2(1-e^{-\sigma}). $$
In particular, these quantities are fractions of $\bar{w}_L$, $\bar{w}_L^2$, respectively, depending only on the ratio $\sigma=\frac{\lambda}{\alpha\mu}$  between the variance $\lambda$ of the random fluctuation $\eta$ and the portion $\alpha\mu$ of the maximum rate $\mu$ of variation in the number of tickets purchased by a gambler in a single round. If
$$ \sigma>2\log{\frac{\bar{w}_L}{\bar w}} $$
then the asymptotic mean $m_\infty$ is lower than the fixed ideal number $\bar{w}$ of tickets to be purchased in each round. This identifies a population of gamblers capable of not being too deeply involved in the jackpot game.

\begin{figure}[!t]
\centering
\includegraphics[scale=0.4]{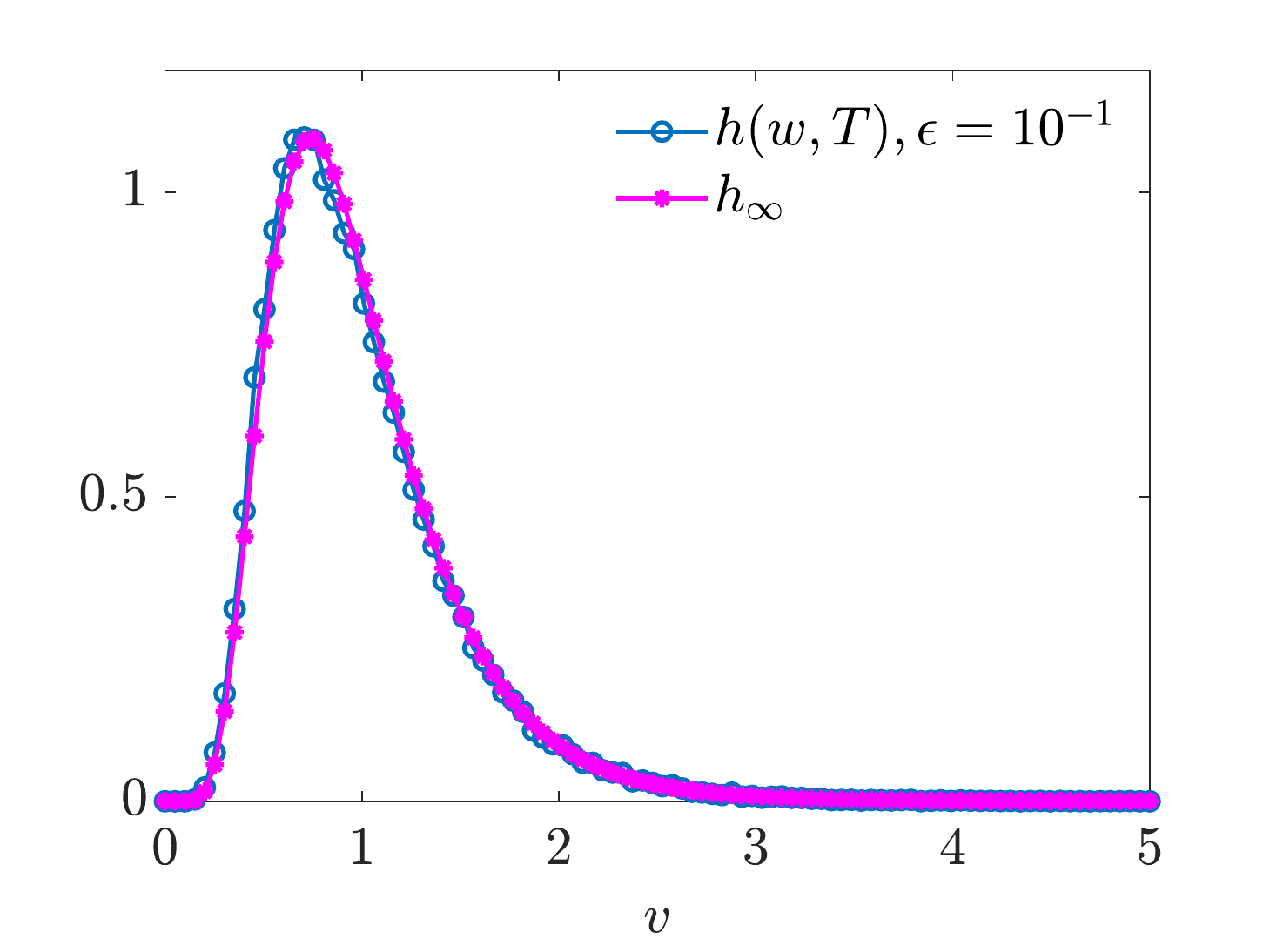}
\includegraphics[scale=0.4]{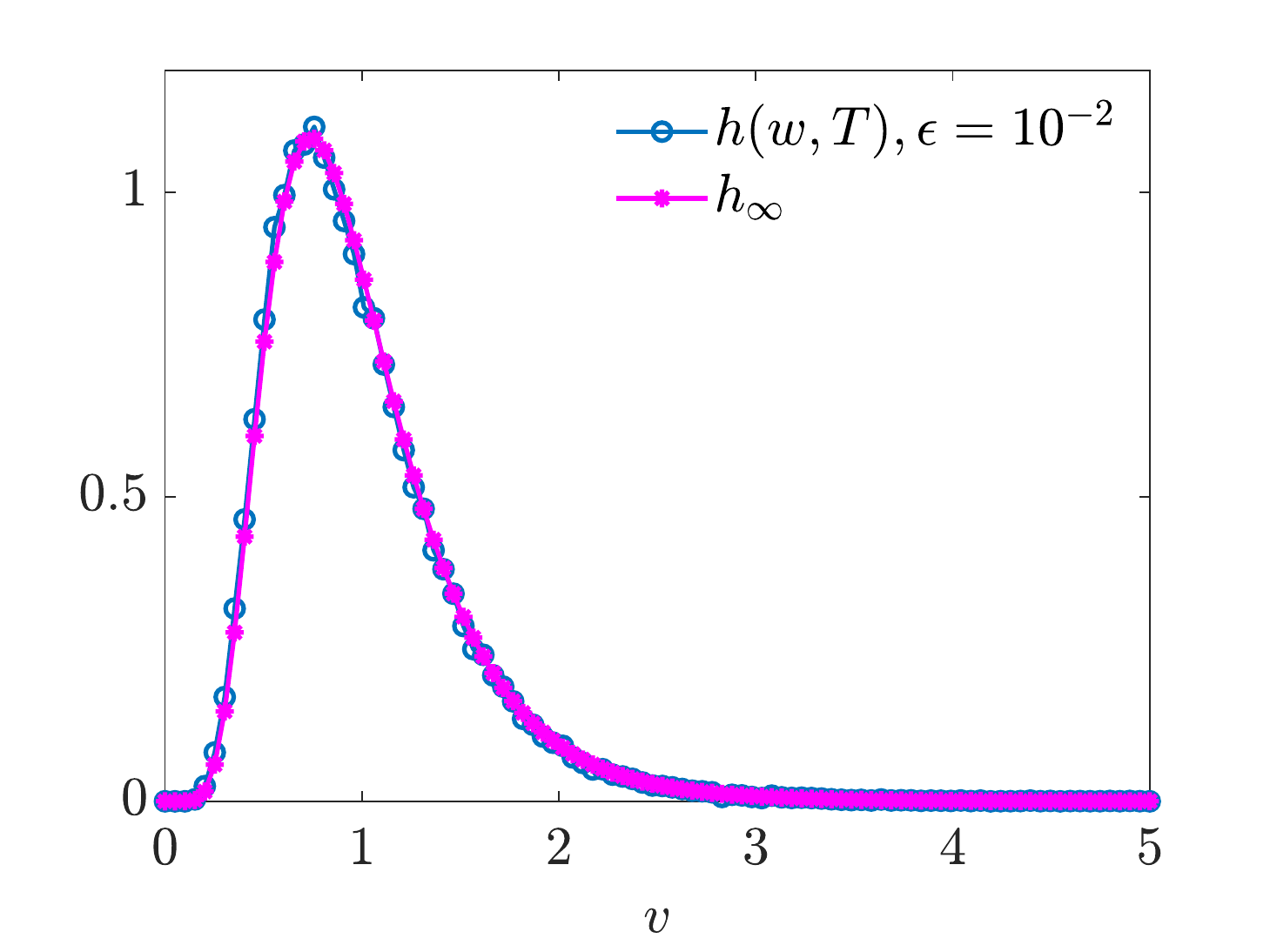}
\caption{Comparison of~\eqref{eq:hinf_log} with the numerically computed large time solution (at the computational time $T=10$) to the Boltzmann-type equation~\eqref{kin-w}. The value function is~\eqref{vf} with $\mu=0.5$, $\alpha=1$. Moreover, the binary interaction is~\eqref{coll} with $\lambda=\frac{1}{10}$ and $w_L=e^\frac{\lambda}{2\mu}$. We considered the quasi-invariant scaling~\eqref{scal} with $\epsilon=10^{-1}$ (left panel) and $\epsilon=10^{-2}$ (right panel).}
\label{fig:lognormal}
\end{figure}

Figure~\ref{fig:lognormal} shows that the asymptotic profile~\eqref{eq:hinf_log} describes excellently the large time distribution of the Boltzmann-type equation~\eqref{kin-w} in the quasi-invariant regime (i.e., $\epsilon$ small in~\eqref{scal}). The solution to~\eqref{kin-w} has been obtained numerically via a standard Monte Carlo method.

\begin{remark}
As shown in~\cite{DT19}, using the value function~\eqref{dif-vf} in place of~\eqref{vf} yields a skewed steady state distribution in the class of the generalised gamma densities. Such densities share most of the properties of the lognormal density and, as it happens in the problem of the alcohol consumption, might provide a better correspondence with the empirically observed profiles if they are used to fit the number of tickets purchased by the gamblers. In any case, the main aspect of the steady state, namely its rapid decay at infinity due to a \textit{slim tail}, remains unchanged.
\end{remark}

\section{Numerical tests}
\label{Numerics}
In this section, we provide numerical insights into the various models discussed before, resorting to direct Monte Carlo methods for collisional kinetic equations and to the recent structure preserving methods for Fokker-Planck equations. For a comprehensive presentation of these numerical methods, the interested reader is addressed to~\cite{Dimarco2014,pareschi2013BOOK,Pareschi2018a,Pareschi2018}.

We begin by integrating the multiple-interaction Boltzmann-type model~\eqref{kine1}, so as to assess its equivalence with the linearised model~\eqref{kine2} in the case $N\gg 1$ with $\epsilon N=\kappa>0$, as predicted theoretically in Section~\ref{linea1}. Next, we also test numerically the consistency of the Fokker-Planck equation~\eqref{FPmod} with the linearised Boltzmann-type equation~\eqref{kine2}. Subsequently, we investigate the kinetic model with fat tails discussed in Section~\ref{sect:FP.jackpot}. In particular, we evaluate numerically some discrepancies that it presents with the other models.

\subsection{Test 1: the multiple-interaction Boltzmann-type model and its linearised version}
\label{test1}
The multiple-interaction Boltzmann-type equation~\eqref{kine1} can be fruitfully written in strong form, to put in evidence the gain and loss parts of the integral operator:
\begin{align}
    \begin{aligned}[b]
        \partial_t f(x,\,t) &= \frac{1}{\epsilon}\ave*{\int_{\R^{N-1}}\left(\frac{1}{J}\prod_{k=1}^N f(\pr{x}_k,\,t)-\prod_{k=1}^N f(x_k,\,t)\right)\,dx_2\,\dots\,dx_N} \\
        &= \frac{1}{\epsilon}Q^+(f,\,\dots,\,f)(x,\,t)-\dfrac{1}{\epsilon}f(x,\,t),
    \end{aligned}
    \label{eq:kine1_strong}
\end{align}
where $Q^+$ is the gain operator:
$$ Q^+(f,\,\dots,\,f)(x,\,t):=\frac{1}{\epsilon}\ave*{\int_{\R^{N-1}}\prod_{k=1}^{N}\frac{1}{J}f(\pr{x}_k,\,t)\,dx_2\,\dots\,dx_N} $$
and $J$ is the Jacobian of the transformation~\eqref{micro} from the pre-interaction variables $\{\pr{x}_k\}_{k=1}^{N}$ to the post-interaction variables $\{x_k\}_{k=1}^{N}$

We discretise~\eqref{eq:kine1_strong} in time through a forward scheme on the mesh $t^n:=n\Delta{t}$, $\Delta{t}>0$. With the notation $f^n(x):=f(x,\,t^n)$, we obtain the following semi-discrete formulation:
$$ f^{n+1}(x)=\left(1-\frac{\Delta{t}}{\epsilon}\right)f^n(x)+\frac{\Delta{t}}{\epsilon}Q^+(f^n,\,\dots,\,f^n)(x). $$
By choosing $\Delta{t}=\epsilon$, the loss part disappears and at each time step only the gain operator $Q^+$ needs to be computed.

We recall that the multiple-interaction microscopic dynamics are given by~\eqref{micro}. In particular, motivated by the results of Section~\ref{Behavior}, we choose the $Y_k$'s as independent and identically distributed random variables with lognormal probability density:
\begin{equation}
    \Phi(y)=\frac1{\sqrt{4\pi}y}\exp\left(-\frac{{(\log{y}+1)}^2}{2}\right).
    \label{eq:sample_refilling}
\end{equation}
A comparison with~\eqref{eq:hinf_log} shows that this corresponds to $\sigma=2$ and $\bar{w}_L=e$, so that $M=\ave{Y_k}=1$ for all $k$.

Parallelly, we consider the linearised Boltzmann-type equation~\eqref{kine2}, which we have shown to be formally equivalent to the multiple-interaction model for a large number of gamblers $N$. The semi-discrete in time formulation of the linearised model reads 
$$ f^{n+1}(x)=\left(1-\frac{\Delta{t}}{\epsilon}\right)f^n(x)+\frac{\Delta{t}}{\epsilon}\ave*{\int_{\R_+}\frac{1}{J}f^n(\pr{x})\,dx}, $$
where now the microscopic dynamics are given by~\eqref{eq:micro.lin-refill} with $\kappa=\epsilon N$ and $Y\sim \Phi(y)$ like before, cf.~\eqref{eq:sample_refilling}.

In both cases, we solve the interaction dynamics by a Monte Carlo scheme, considering a random sample of $10^6$ particles with initial uniform distribution in the interval $[0,\,2]$, thus $f_0(x):=f(x,\,0)=\frac{1}{2}\mathbb{1}_{[0,\,2]}(x)$, where $\mathbb{1}$ denotes the characteristic function.

\begin{figure}[!t]
\centering
\subfigure[$t=0.1$]{\includegraphics[scale=0.32]{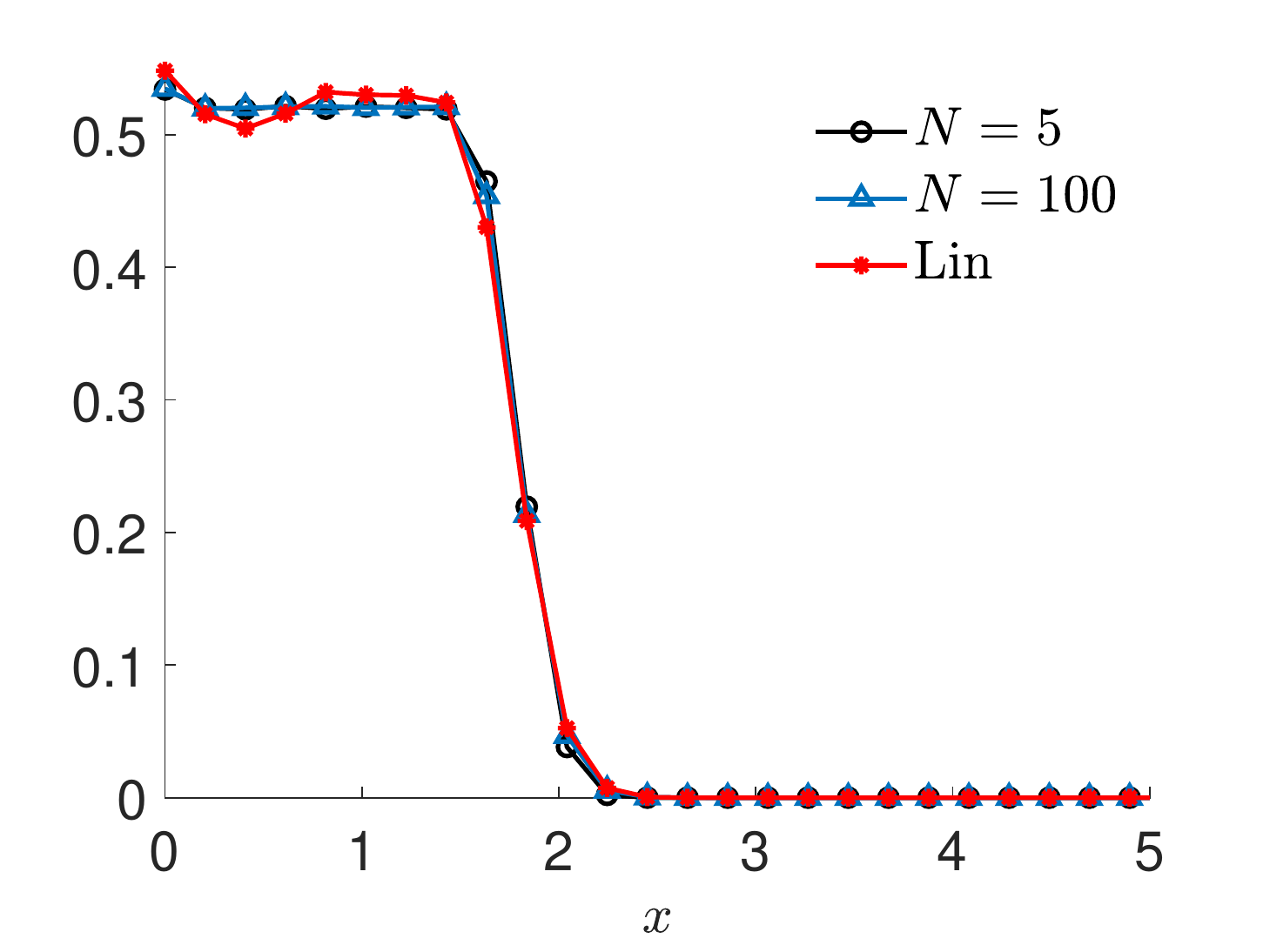}}
\subfigure[$t=1$]{\includegraphics[scale=0.32]{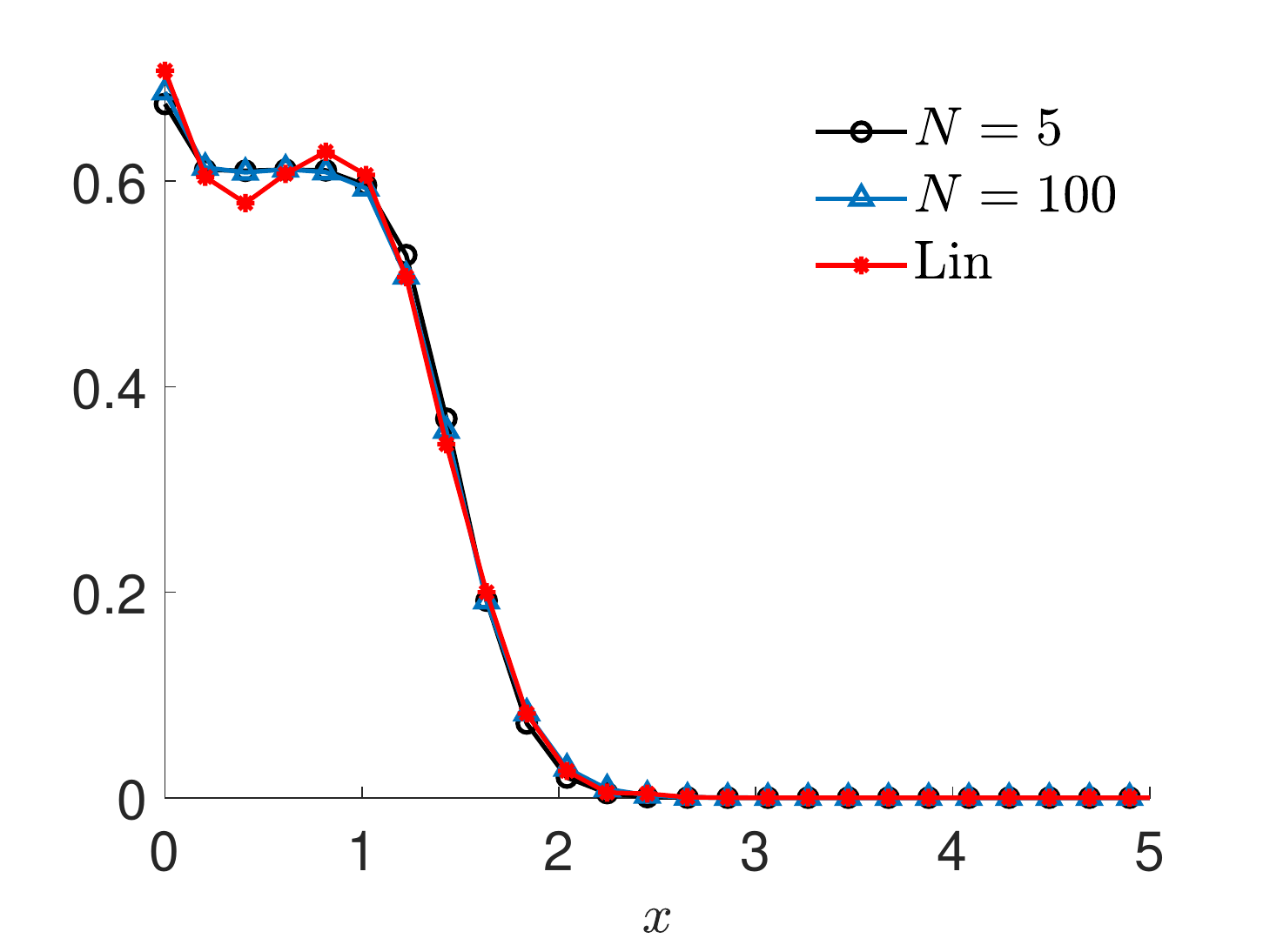}}
\subfigure[$t=2$]{\includegraphics[scale=0.32]{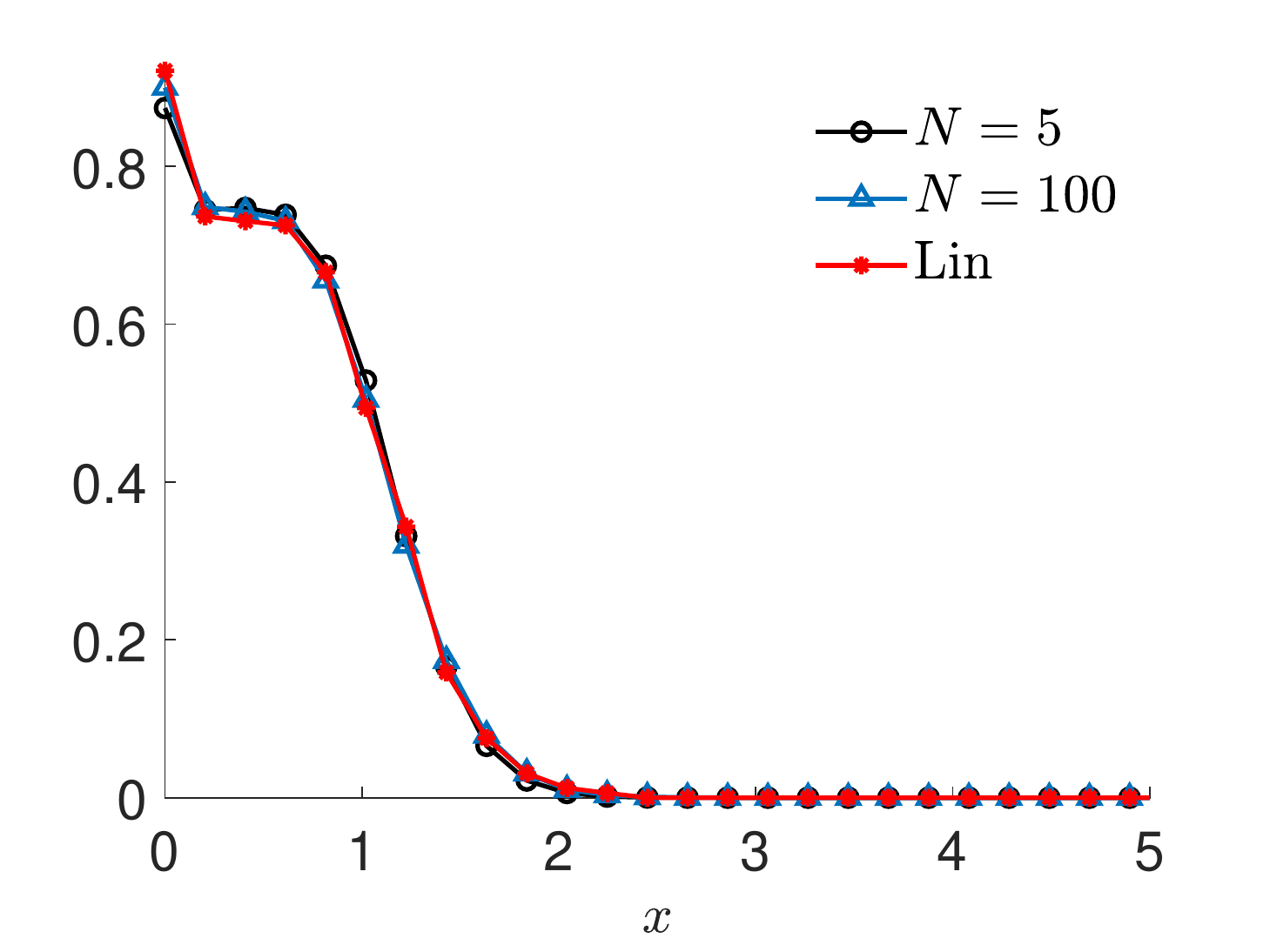}}
\caption{\textbf{Test 1 -- Without refilling}. Evolution of the multiple-interaction Boltzmann-type model with either $N=5$ gamblers (empty circular markers) or $N=100$ gamblers (triangular markers) and of its linearised version (filled circular markers) in the time interval $[0,\,2]$ for $\delta=0.2$, $\beta=0$. We considered $\kappa=0.1$.}
\label{fig:Ncoll_lin}
\end{figure}

\begin{figure}[!t]
\centering
\subfigure[$t=1$]{\includegraphics[scale=0.32]{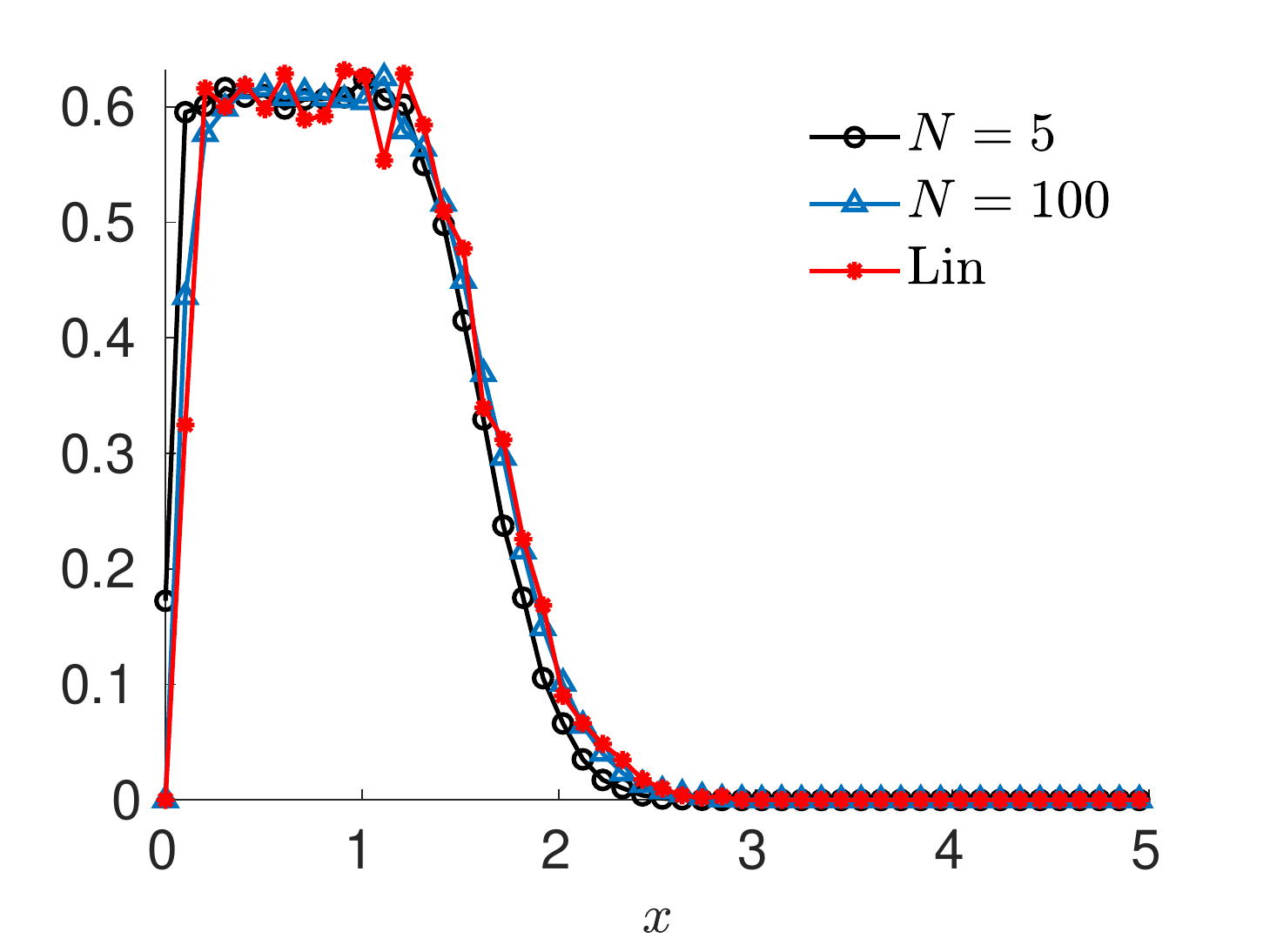}}
\subfigure[$t=5$]{\includegraphics[scale=0.32]{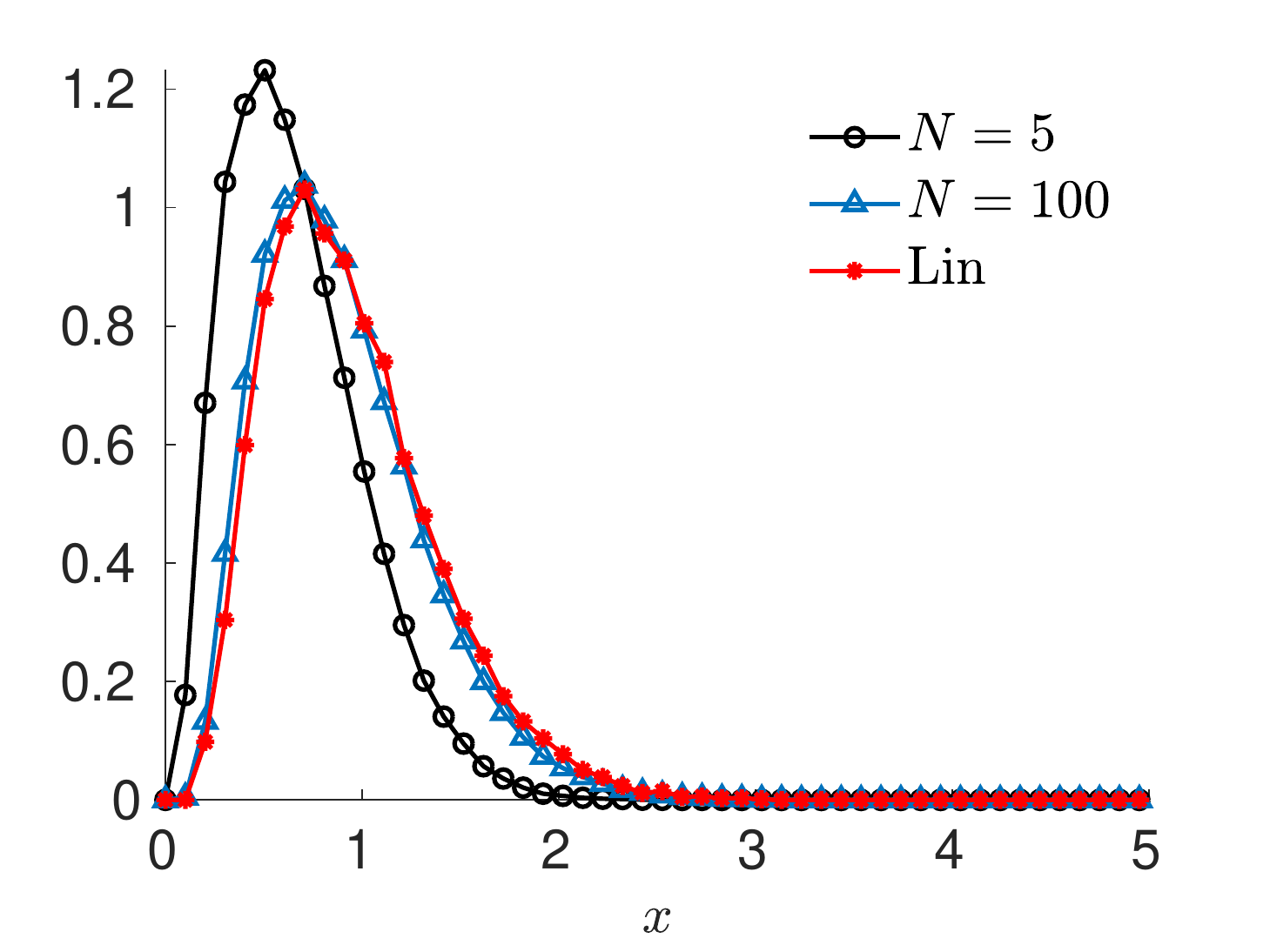}} 
\subfigure[$t=25$]{\includegraphics[scale=0.32]{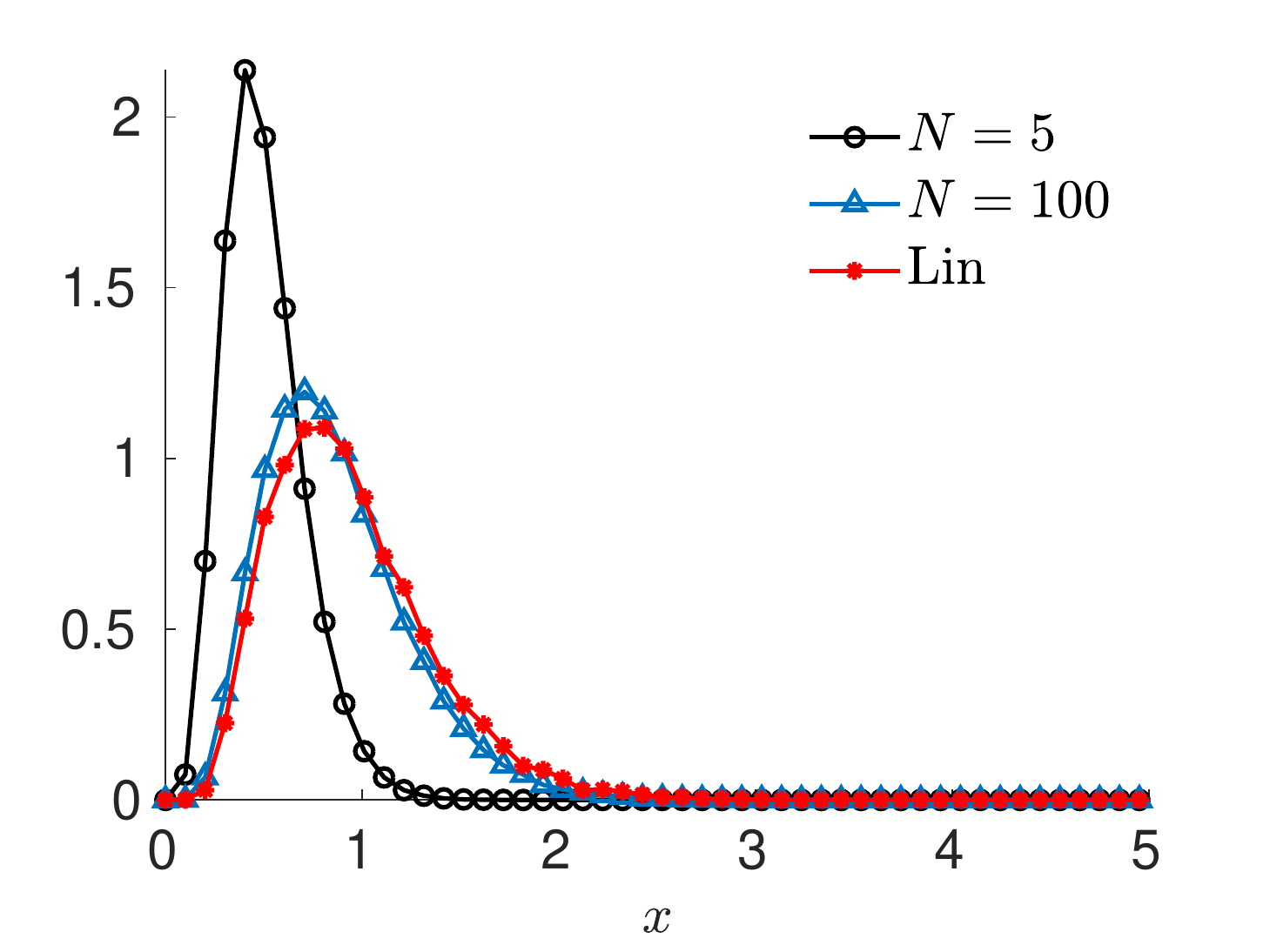}}
\caption{\textbf{Test 1 -- With refilling}. Evolution of the multiple-interaction Boltzmann-type model with either $N=5$ gamblers (empty circular markers) or $N=100$ gamblers (triangular markers) and of its linearised version (filled circular markers) in the time interval $[0,\,25]$ for $\delta=\beta=0.2$ (lognormal refilling sampled from~\eqref{eq:sample_refilling}). We considered $\kappa=0.1$.}
\label{fig:Ncoll_lin_beta02}
\end{figure}

In Figure~\ref{fig:Ncoll_lin}, we compare the evolutions of the two  models in the time interval $t\in [0,\,2]$ for $\delta=0.2$, $\beta=0$ in~\eqref{micro},~\eqref{eq:micro.lin-refill}, cf. also~\eqref{micro1}, i.e., in particular, with no refilling. In Figure~\ref{fig:Ncoll_lin_beta02}, we perform the same test in the larger time interval $t\in [0,\,25]$ for $\delta=\beta=0.2$, i.e. by including also the refilling. In both cases, we clearly see that, if $N$ is sufficiently large, the linearised model is able to catch the multiple-interaction dynamics at each time, whereas differences can be observed if $N$ is relatively small.

\begin{figure}[!t]
\centering
\includegraphics[scale=0.5]{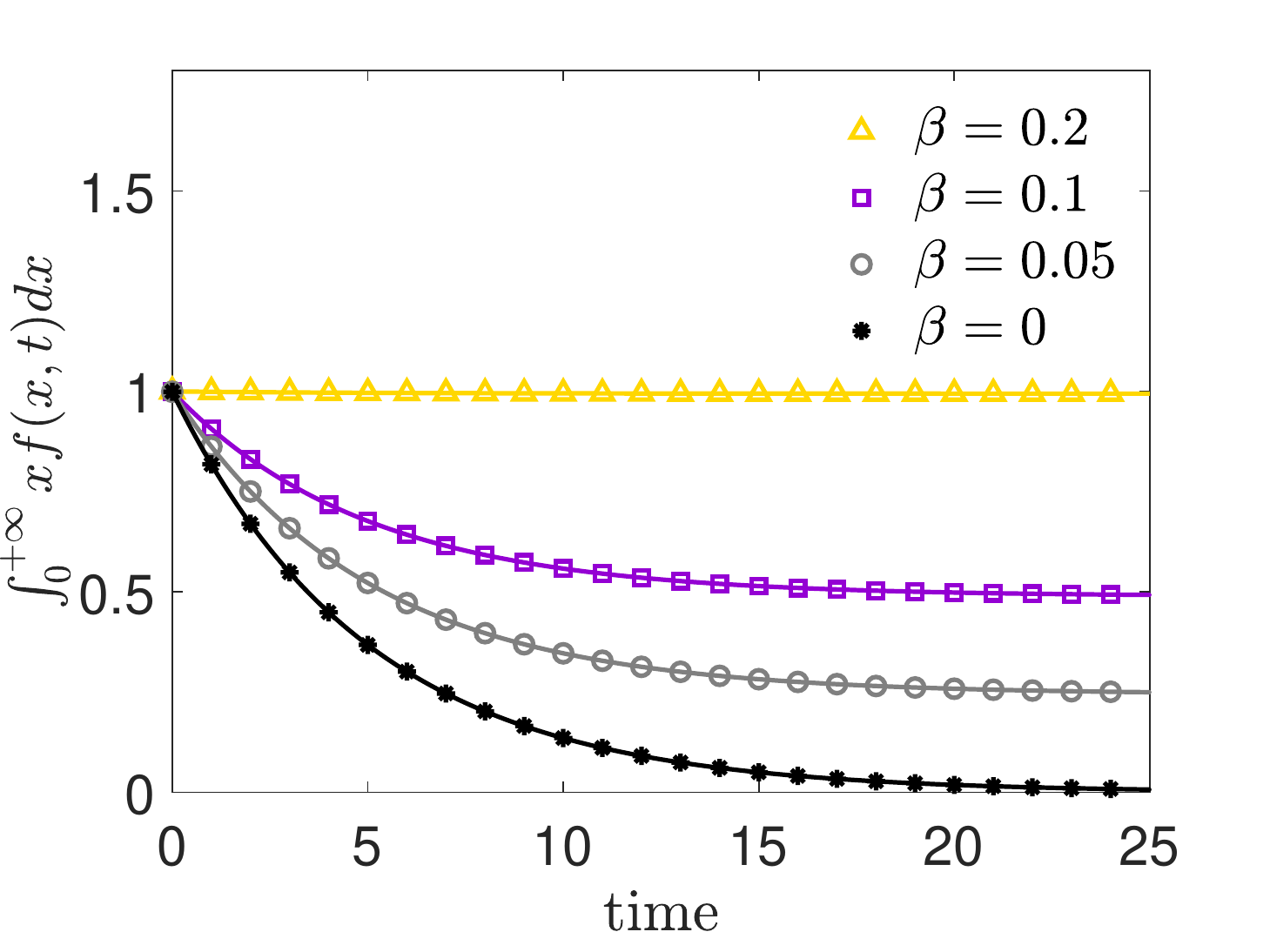}
\caption{\textbf{Test 1 -- Evolution of the mean}. Evolution of the mean number of tickets $m(t)$ in the time interval $[0,\,25]$ for $\delta=0.2$, $\kappa=0.1$ and several choices of $\beta$.}
\label{fig:mean_comp}
\end{figure}

Moreover, in the linearised model, we know that the mean number of tickets owned by the gamblers during the jackpot game is given by~\eqref{expl}. In Figure~\ref{fig:mean_comp}, we show instead the time evolution of the mean of the solution to the multiple-interaction Boltzmann-type model for several choices of the refilling parameter $\beta$. We observe a good agreement with the theoretical results and, in particular, we see that the mean value tends indeed asymptotically to $\frac{\beta M}{\delta}$, as expected.

\subsection{Test 2. Fokker-Planck approximation for large~\texorpdfstring{$\boldsymbol{N}$}{}}
\label{sect:test_2}
In the case $\epsilon,\,\kappa\ll 1$, the interactions~\eqref{eq:micro.lin-refill} are quasi-invariant, hence the linearised Boltzmann-type model~\eqref{kine2} is well described by the Fokker-Planck equation~\eqref{FPmod}. In the case of a constant mean value $m(t)\equiv m_0=\frac{\beta M}{\delta}$ of the number of tickets owned by the gamblers, the steady distribution is the gamma probability density function~\eqref{eq:gamma}. In this section, we compare numerically the large time distributions produced by either the multiple-interaction Boltzmann-type model~\eqref{kine1} or the linearised Boltzmann-type model~\eqref{kine2} with~\eqref{eq:gamma}.

Like before, we consider a uniform initial distribution $f_0(x)$ in the interval $[0,\,2]$ and moreover a random variable $Y$ lognormally distributed according to~\eqref{eq:sample_refilling}, thus in particular with mean $M=1$. We also set $\beta=\delta=0.2$ in the microscopic interactions~\eqref{micro},~\eqref{eq:micro.lin-refill}, so that the mean value of the ticket distribution is always $m_0=\frac{\beta M}{\delta}=1$, consistently with the Fokker-Planck regime in which we are able to compute explicitly the steady distribution~\eqref{eq:gamma}.

\begin{figure}[!t]
\centering
\subfigure[]{\includegraphics[scale=0.5]{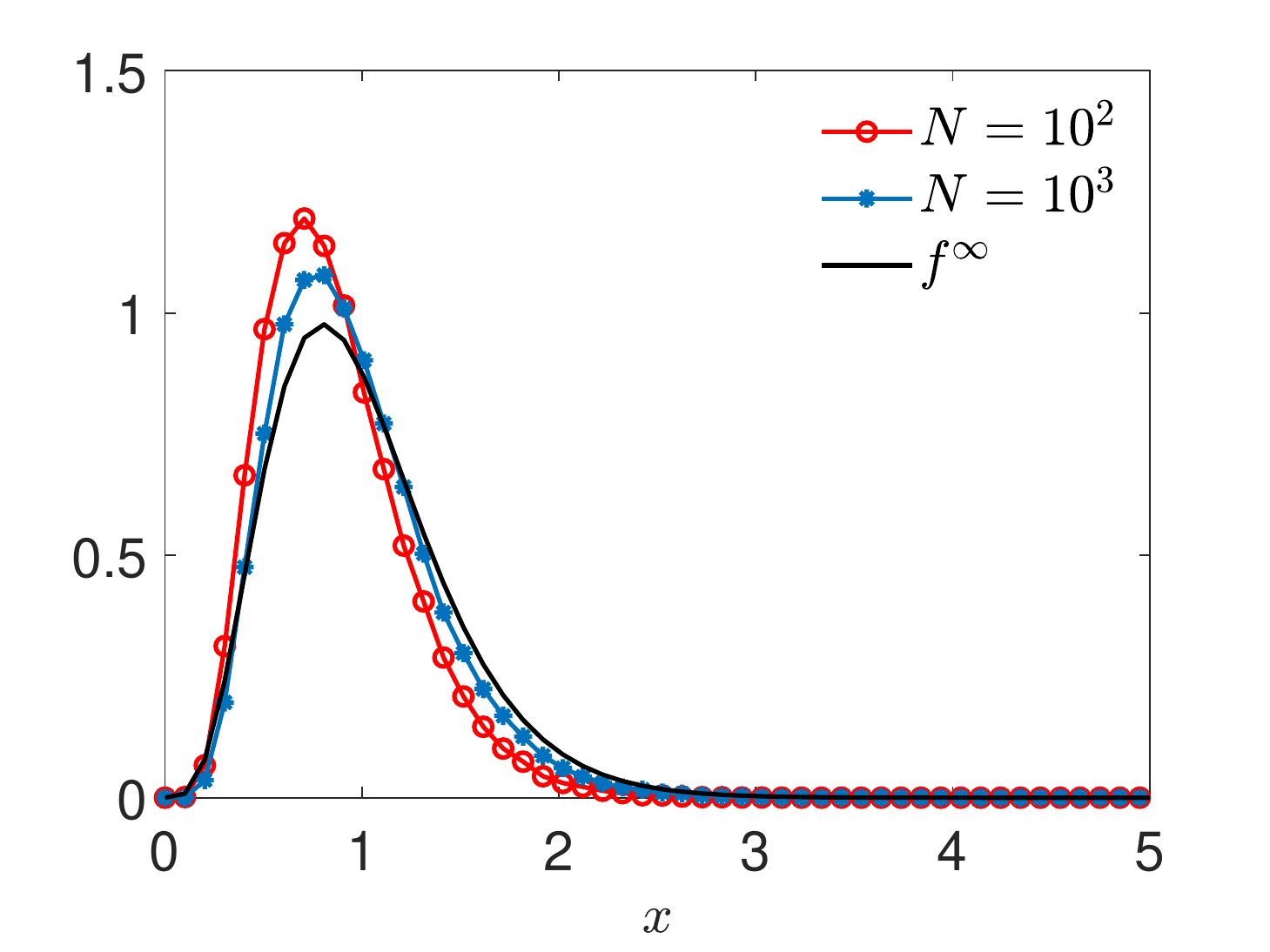}}
\subfigure[]{\includegraphics[scale=0.5]{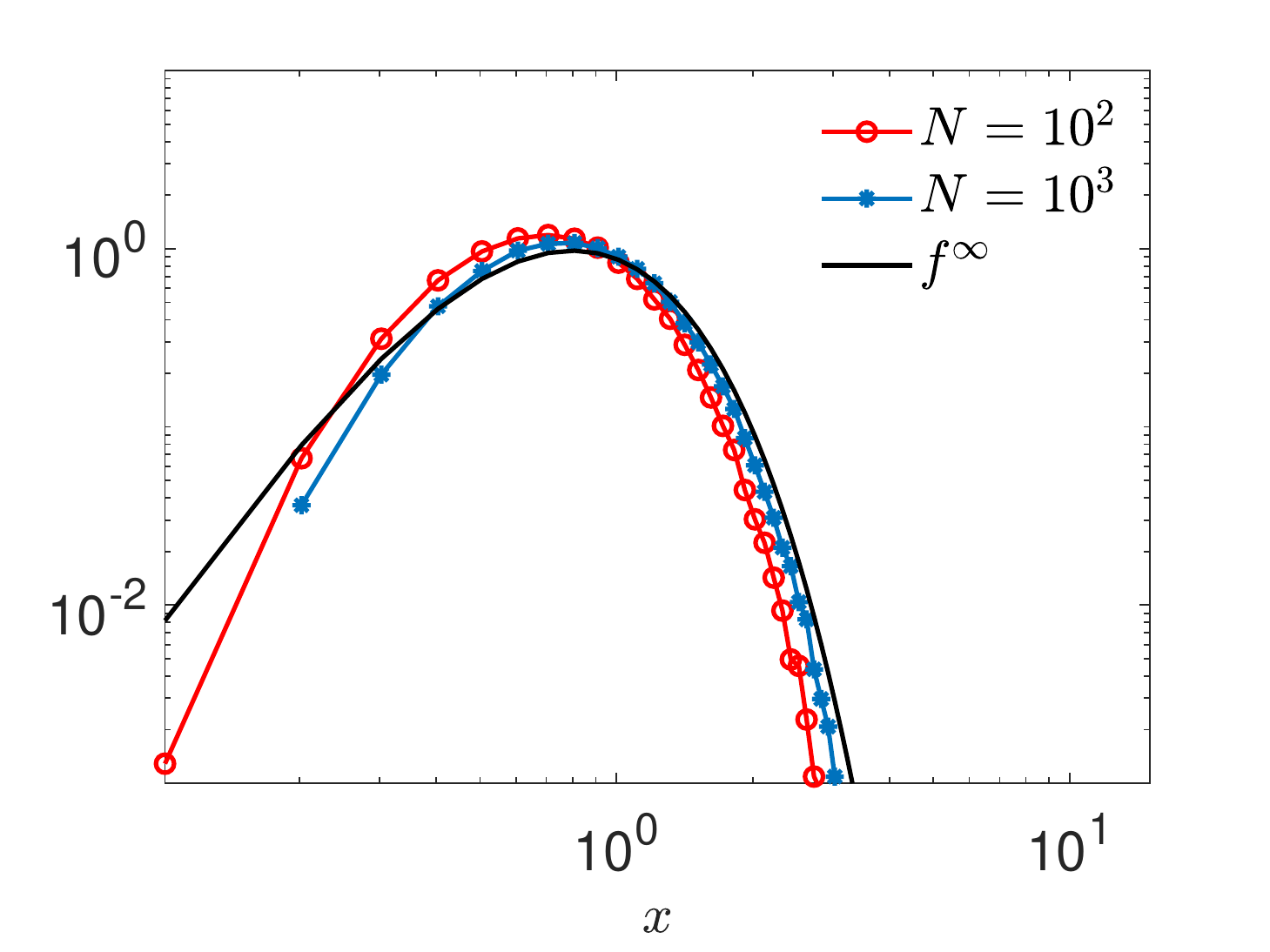}} \\
\subfigure[]{\includegraphics[scale=0.5]{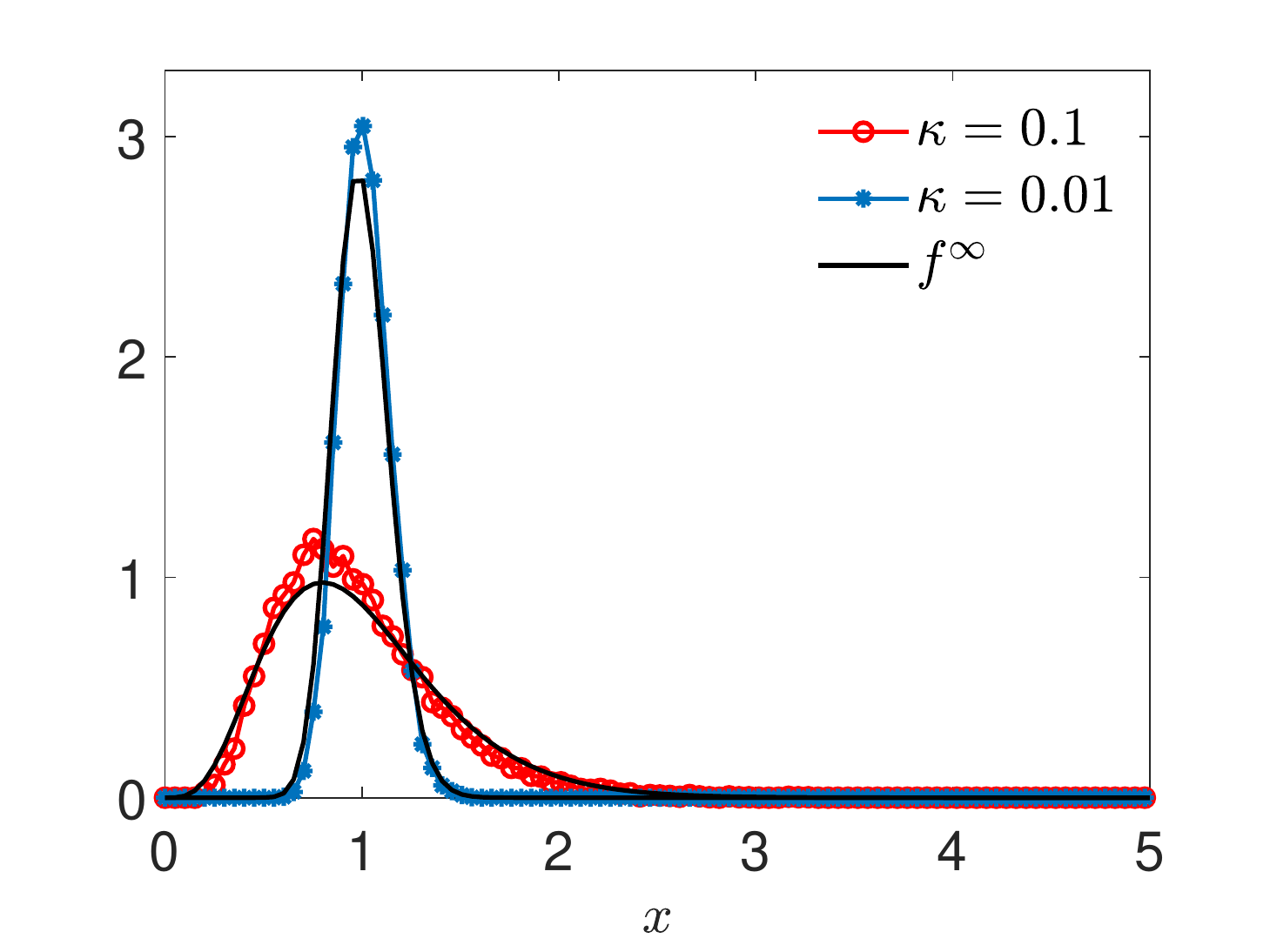}}
\subfigure[]{\includegraphics[scale=0.5]{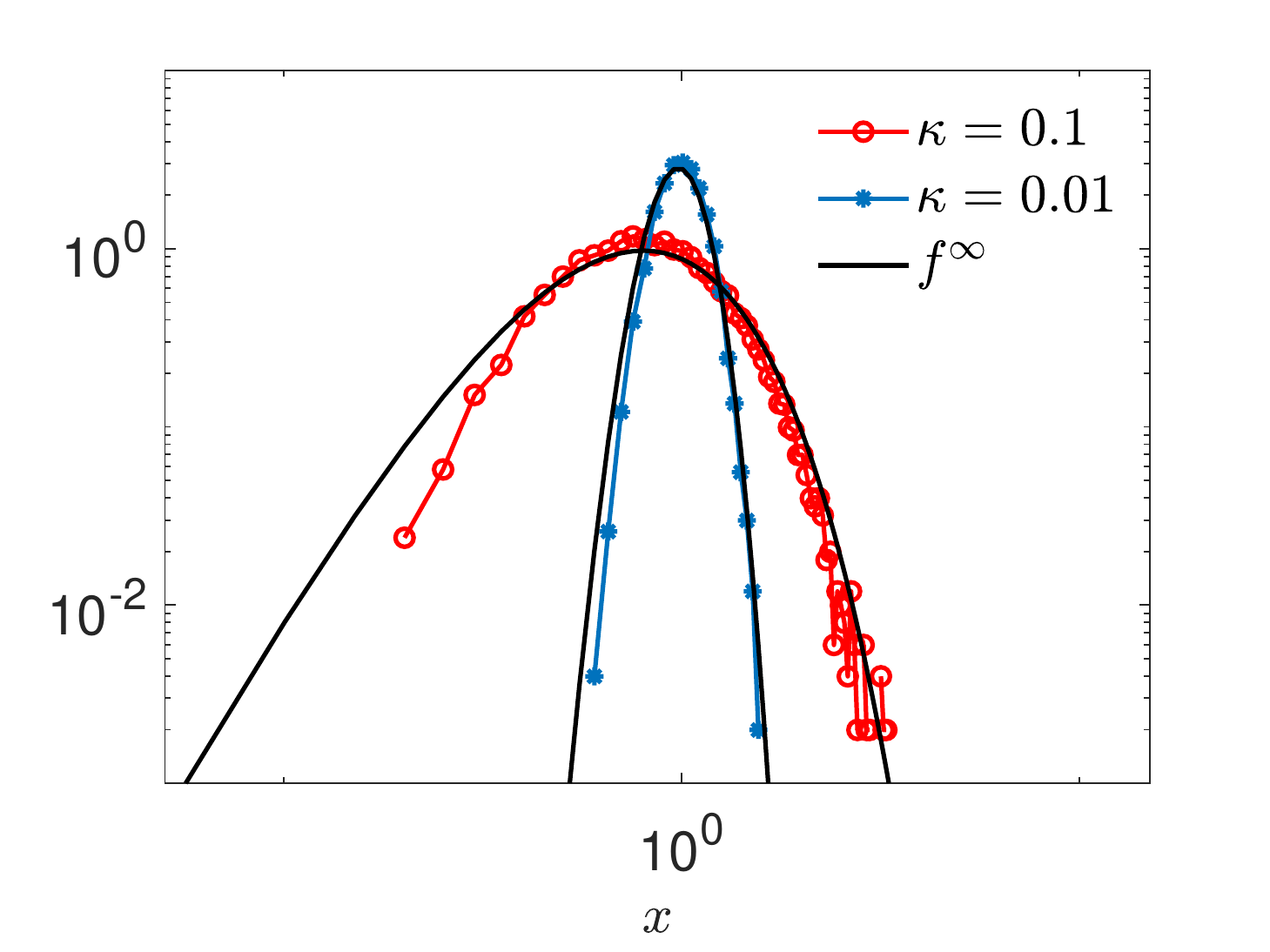}}
\caption{\textbf{Test 2}. Top row: (a) Comparison of the steady distribution of the multiple-interaction Boltzmann-type model~\eqref{kine1} with the Fokker-Planck asymptotic distribution~\eqref{eq:gamma} (solid line) for $N=10^2$ (empty circular markers), $N=10^3$ (filled circular markers) and fixed $\kappa=0.1$. (b) Log-log plot of (a). Bottom row: (c) Comparison of the steady distribution of the linearised Boltzmann-type model~\eqref{kine2} with the Fokker-Planck asymptotic distribution~\eqref{eq:gamma} (solid line) for $\kappa=0.1$ (empty circular markers) and $\kappa=0.01$ (filled circular markers). (d) Log-log plot of (c).}
\label{fig:asymp}
\end{figure}

In Figure~\ref{fig:asymp}(a), we compare the large time distribution of the multiple-interaction Boltzmann-type model for an increasing number of gamblers participating in each round of the jackpot game ($N=10^2$, $N=10^3$, respectively) with the asymptotic gamma probability density~\eqref{eq:gamma} computed from the Fokker-Planck equation. We clearly see that, for $N$ large enough, the Fokker-Planck steady solution provides a good approximation of the equilibrium distribution of the real multiple-interaction model. In Figure~\ref{fig:asymp}(b), we show the log-log plot of the same distributions, which allows us to appreciate that, in particular, the Fokker-Planck solution reproduces correctly the tail of the equilibrium distribution of the multiple-interaction model, thereby confirming that no fat tails have to be expected in the distribution of the tickets owned by the gamblers.

In Figure~\ref{fig:asymp}(c), we compare instead the large time distribution of the linearised Boltzmann-type model with the asymptotic gamma probability density~\eqref{eq:gamma} for decreasing values of $\kappa$ ($\kappa=0.1$, $\kappa=0.01$, respectively). In Figure~\ref{fig:asymp}(d), we show the log-log plot of the same distributions to stress, in particular, the goodness of the approximation of the tail provided by~\eqref{eq:gamma}.

\subsection{Test 3. The fat tail case}
In Section~\ref{sect:FP.jackpot}, we derived the alternative linear Boltzmann-type model~\eqref{new1}-\eqref{kine-full}, which preserves some of the main macroscopic properties of the original multiple-interaction model~\eqref{micro}-\eqref{kine1}. In particular, it accounts for the right evolution of the first and second moment of the distribution function.

\begin{figure}[!t]
\centering
\subfigure[]{\includegraphics[scale=0.5]{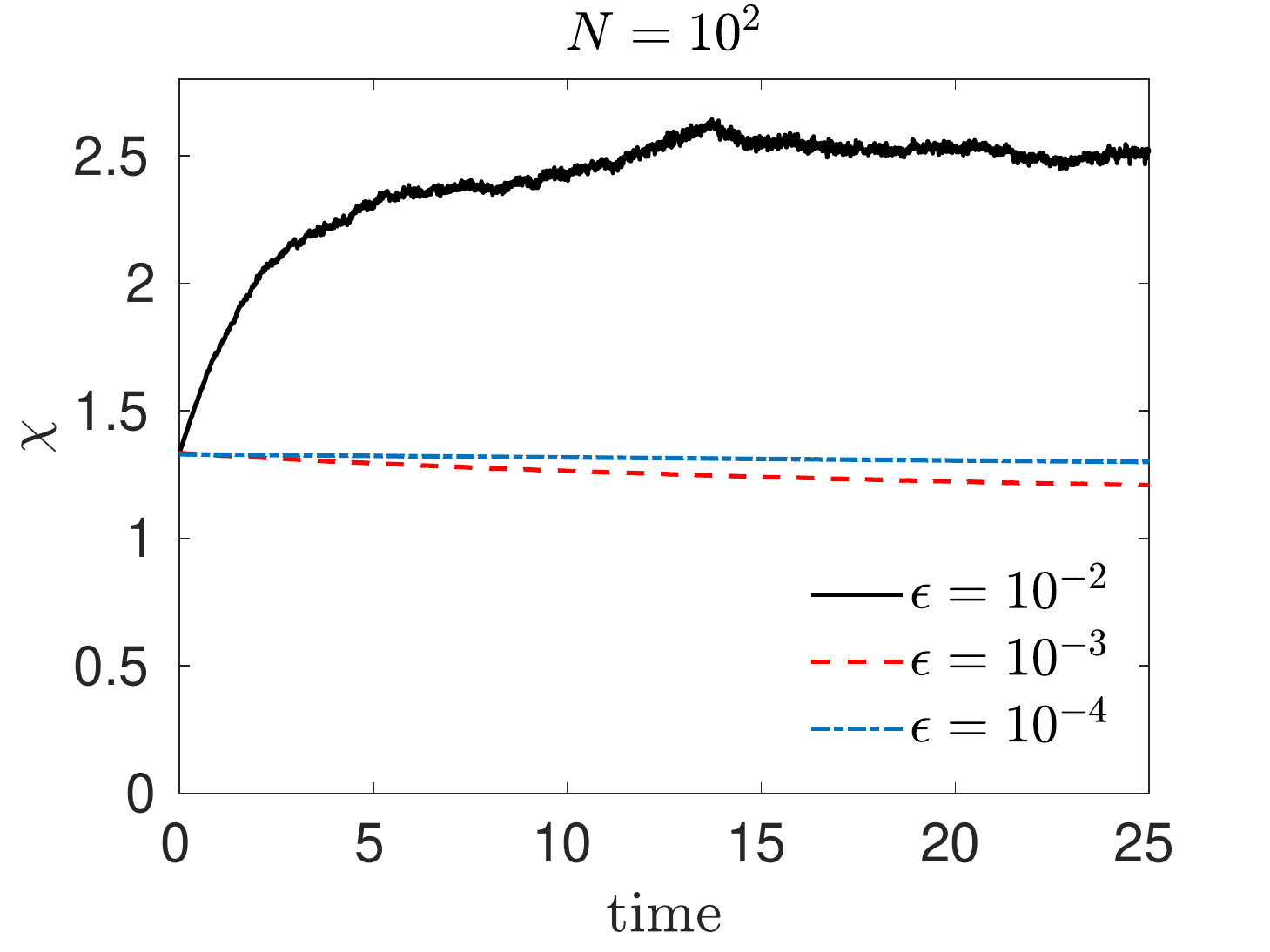}}
\subfigure[]{\includegraphics[scale=0.5]{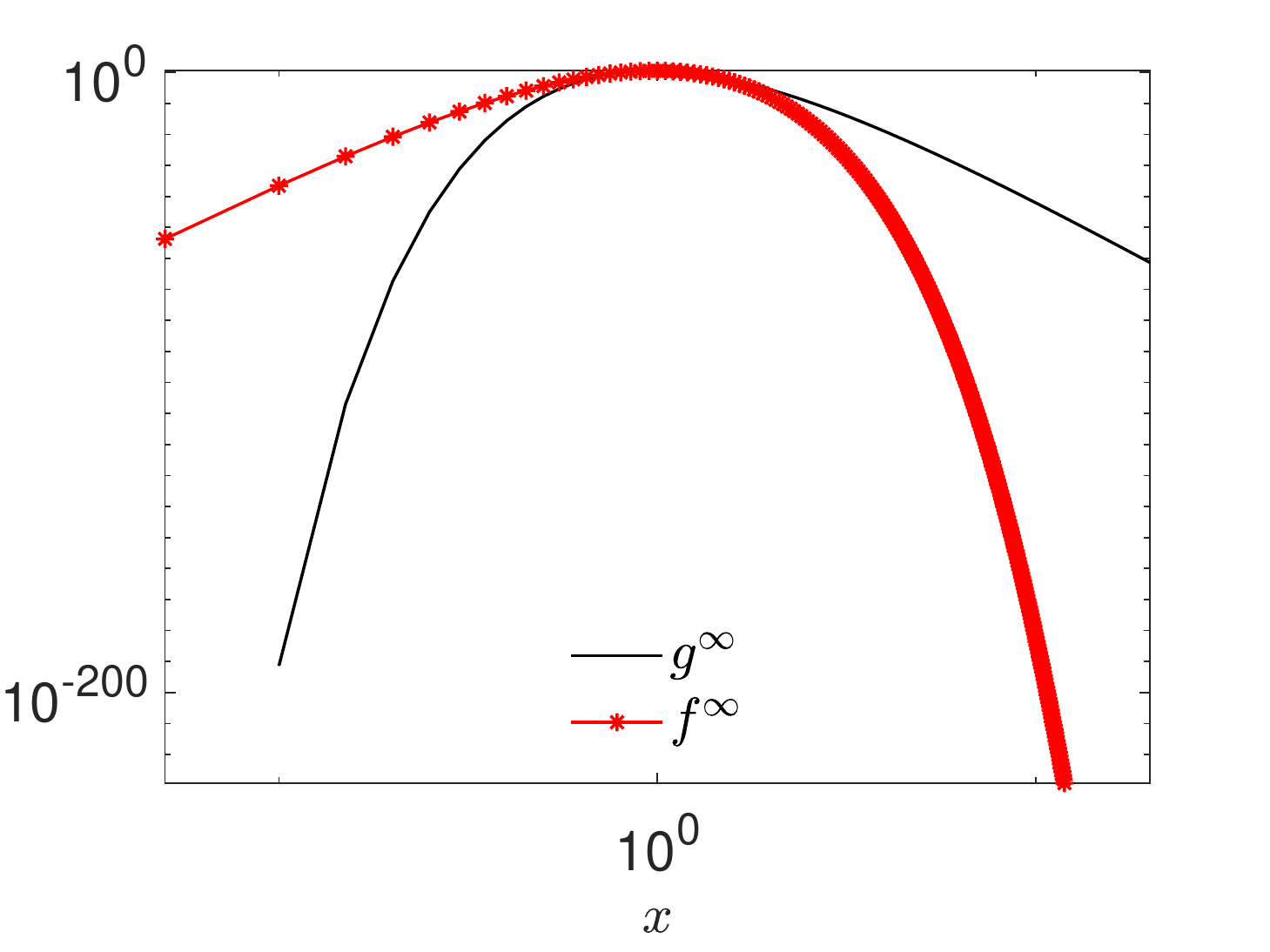}}
\caption{\textbf{Test 3}. (a) Estimate of the approximate collision invariant $\chi$, cf.~\eqref{eq:chi}. (b) Log-log plot of the distributions~\eqref{eq:gamma},~\eqref{eq:invgamma}.}
\label{fig:chi}
\end{figure}

We grounded such a derivation on the consideration that, for $N$ large and $\epsilon$ small, the quantity $\chi$ defined in~\eqref{eq:chi} may be treated approximately as a collision invariant of the $N$-gambler dynamics. In Figure~\ref{fig:chi}(a), we test numerically this assumption by taking $N=10^2$ and some values of the scaling parameter $\epsilon$ decreasing from $10^{-2}$ to $10^{-4}$. In particular, since $\chi$ depends actually on the evolving microscopic states $x_1,\,\dots,\,x_N$ of the agents, we plot the time evolution of $\chi$ for $t\in [0,\,25]$. Such a time evolution is computed with the Monte Carlo method described in Section~\ref{test1}, starting from an initial sample of $S=10^6$ particles. Therefore, we get $N=10^2$ sub-samples of $S/N=10^4$ particles, each of which produces a Monte Carlo estimate of the time trend of $\chi$. Out of these samples, we compute finally the average time trend of $\chi$, namely each of the curves plotted in Figure~\ref{fig:chi}(a). Consistently with our theoretical findings, we observe that, for $\epsilon$ small enough, $\chi$ may be actually regarded as a collision invariant.

In the quasi-invariant limit, the solution to the linear Boltzmann-type model~\eqref{new1}-\eqref{kine-full} has been shown to approach that of the Fokker-Planck equation~\eqref{FP3}. Its explicitly computable steady state is the inverse gamma probability density~\eqref{eq:invgamma}, which, unlike the equilibrium distribution~\eqref{eq:gamma} approximating the trend of the multiple-interaction model for large $N$, exhibits a fat tail. In Figure~\ref{fig:chi}(b), we show the log-log plot of the distributions~\eqref{eq:gamma},~\eqref{eq:invgamma}, which stresses the difference in their tails.

\begin{figure}[!t]
\centering
\subfigure[$t=1$]{\includegraphics[scale=0.32]{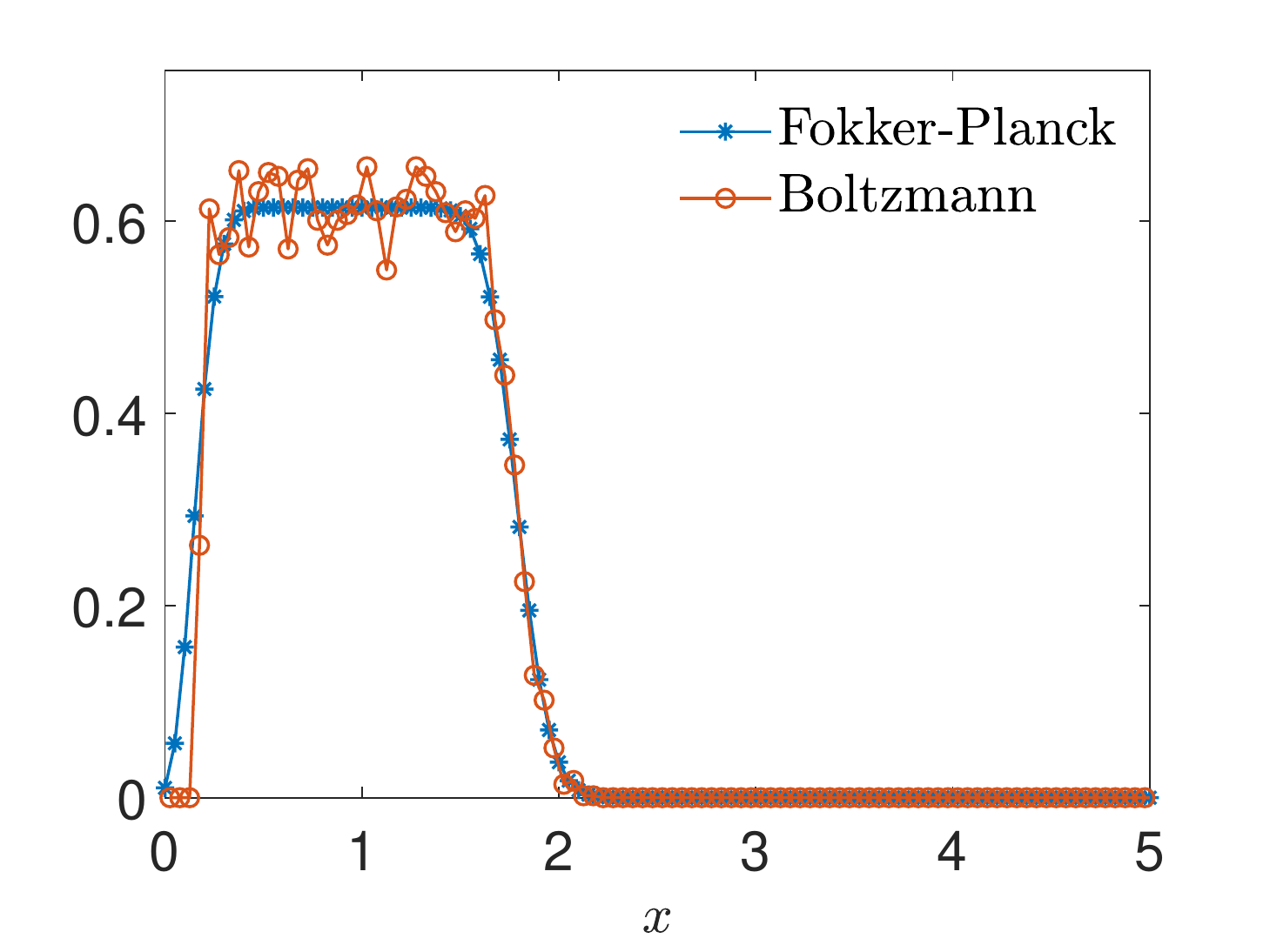}}
\subfigure[$t=5$]{\includegraphics[scale=0.32]{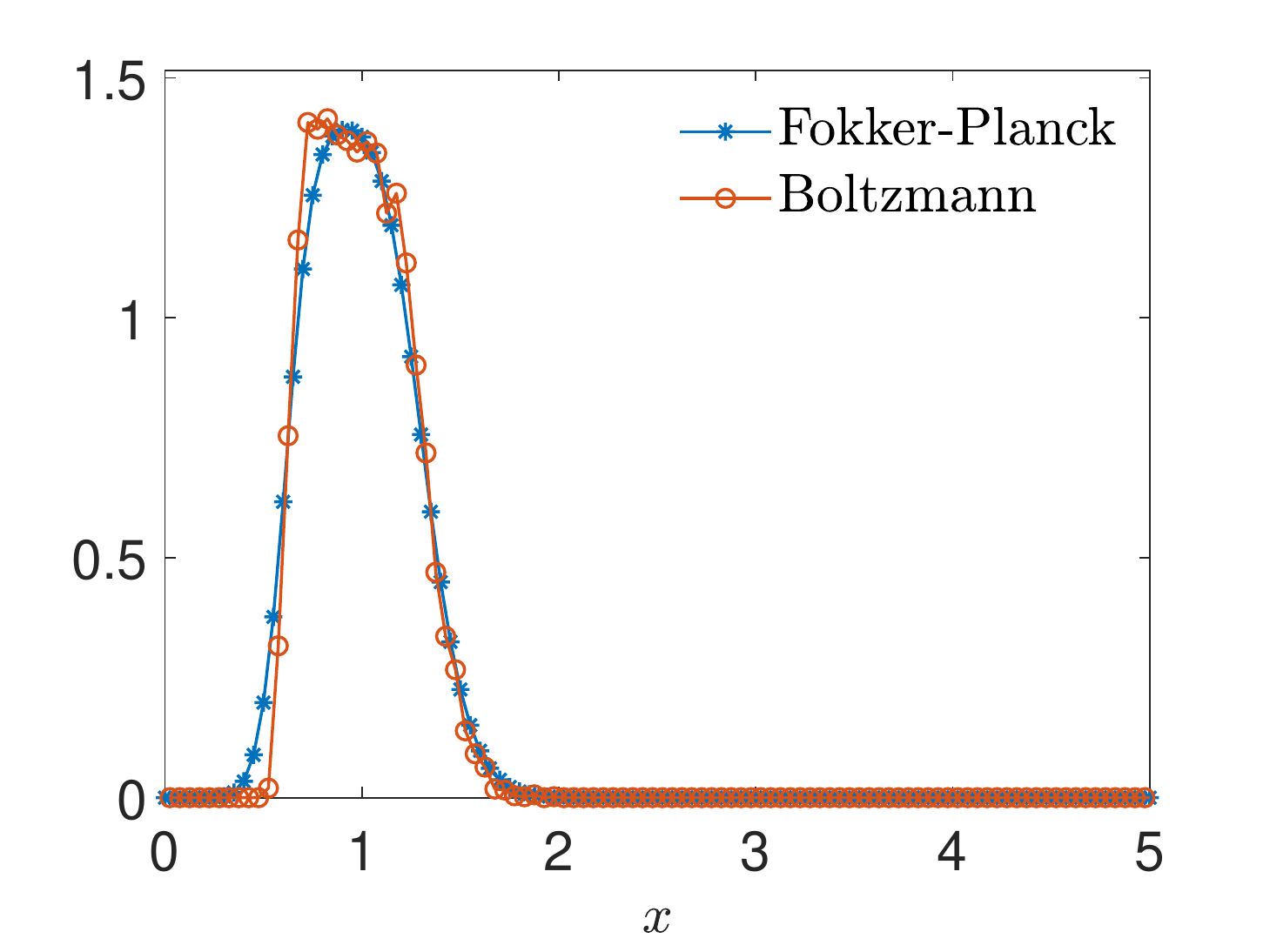}}
\subfigure[$t=25$]{\includegraphics[scale=0.32]{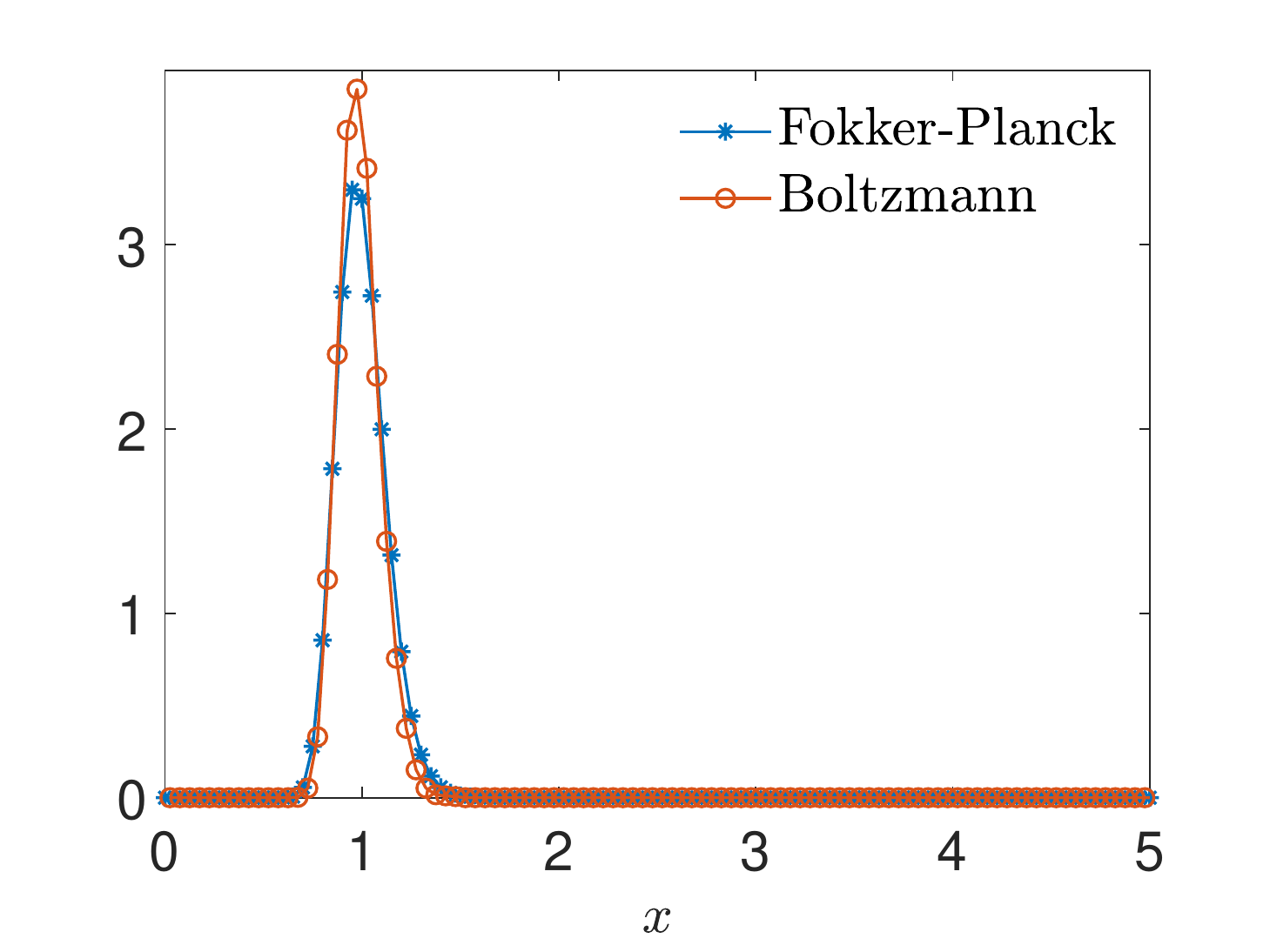}}
\caption{\textbf{Test 3}. Comparison between the time evolutions of the Boltzmann-type model~\eqref{new1},~\eqref{kine-full} (circular markers) and of its Fokker-Planck approximation~\eqref{FP3} (starred markers) in the quasi-invariant regime. The following parameters have been used: $\beta=\delta=0.2$, $M=1$, $\kappa=10^{-2}$. The value of the approximate collision invariant $\chi$ is estimated from the multiple-interaction Boltzmann-type model like in Figure~\ref{fig:chi}.}
\label{fig:newmodel}
\end{figure}

In order to check the consistency of the Fokker-Planck regime described, in the quasi-invariant limit, by~\eqref{FP3} with the Boltzmann-type model~\eqref{new1},~\eqref{kine-full}, in Figure~\ref{fig:newmodel} we show the time evolution of the distribution function $g$ computed with both models for $t\in [0,\,25]$, starting from an initial uniform distribution for $x\in [0,\,2]$. In both cases, we treat $\chi$ as a collision invariant of the $N$-gambler model. Thus, we first computed the value of $\chi$ from model~\eqref{micro},~\eqref{kine1} (with $N=10^4$), then we used it in the binary rules~\eqref{new1}, where $\chi$ determines the values that $\eta_\epsilon$ can take, and in the diffusion coefficient $\tilde{\sigma}$ of~\eqref{FP3}. From Figure~\ref{fig:newmodel}, we see that the two models remain close to each other at every time and approach the same steady distribution for large times, as expected.

\begin{figure}[!t]
\centering
\includegraphics[scale=0.5]{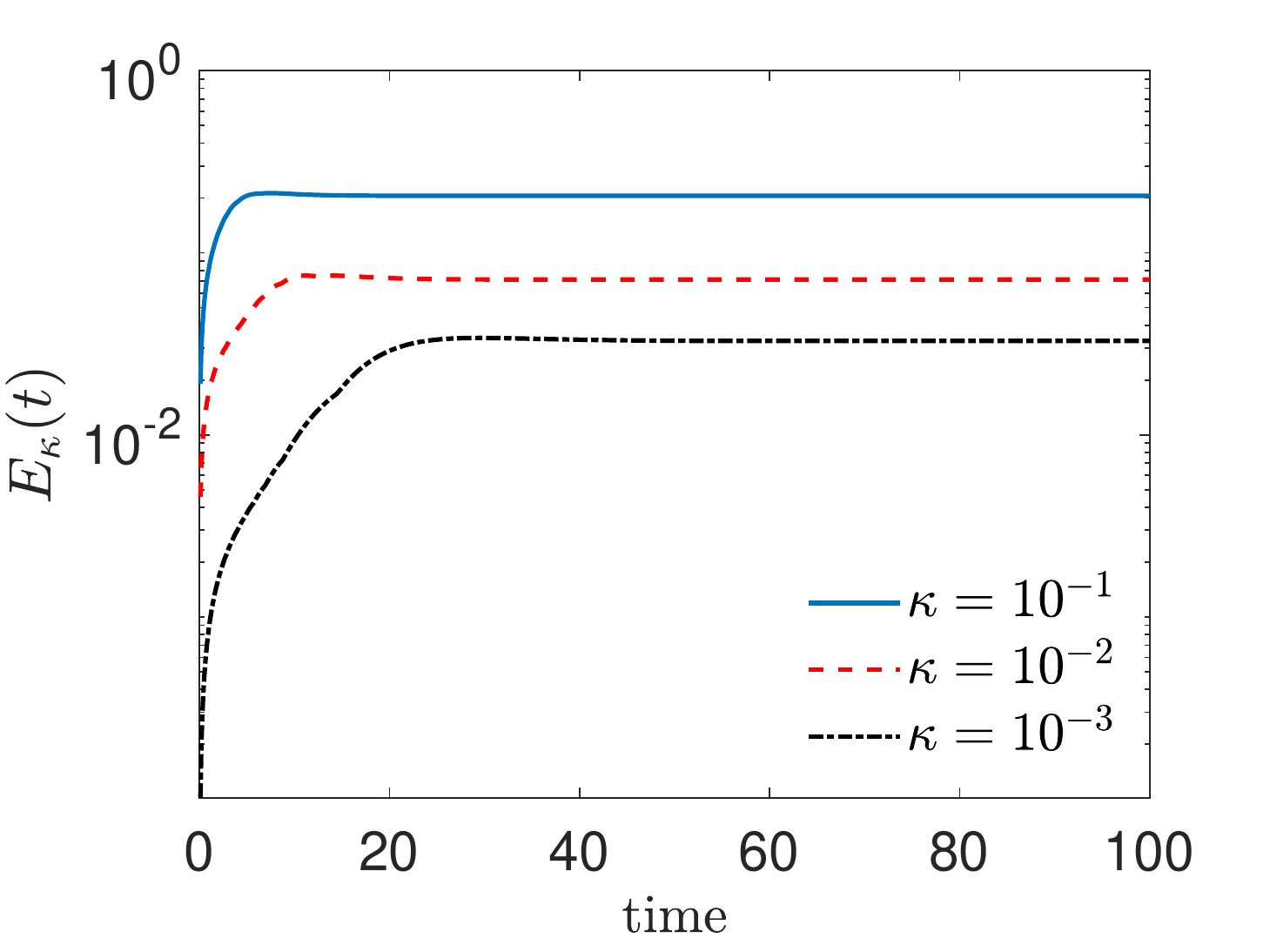}
\caption{\textbf{Test 3}. Relative $L^1$-error~\eqref{eq:Ekappa} between the solutions to the Fokker-Planck equations~\eqref{FPmod} and~\eqref{FP3}. The numerical solution of both models is obtained by means of semi-implicit SP methods over the computational domain $[0,\,10]$ in the $x$ variable, with $\Delta{t}=\Delta{x}= 10/N_x$ and $N_x=401$ nodes.}
\label{fig:err}
\end{figure}

Finally, we quantify the distance between the solution $f$ to the Fokker-Planck equation~\eqref{FPmod}, which reproduces the large time trend of the multiple-interaction Boltzmann-type model~\eqref{micro},~\eqref{kine1}, cf. the previous Test 2 (Section~\ref{sect:test_2}), and the solution $g$ to Fokker-Planck equation~\eqref{FP3}, which describes instead the large time trend of the linear diffusive Boltzmann-type model~\eqref{new1},~\eqref{kine-full}. We consider, in particular, the following relative $L^1$-error
\begin{equation}
    E_\kappa(t):=\int_{\R_+}\frac{\abs*{g(x,\,t)-f(x,\,t)}}{f(x,\,t)}\,dx,
    \label{eq:Ekappa}
\end{equation}
for several values of the constant $\kappa$, cf.~\eqref{bound1}, which appears as a coefficient in both Fokker-Planck equations. In particular, we consider $\kappa=10^{-1},\,10^{-2},\,10^{-3}$ and we take $f(x,\,0)=g(x,\,0)=\frac{1}{2}\mathbb{1}_{[0,\,2]}(x)$ as initial (uniform) distribution. By means of semi-implicit SP methods, we guarantee the positivity and the large time accuracy of the numerical solution to both models. The interested reader is referred to~\cite{Pareschi2018} for further details on this numerical technique). From Figure~\ref{fig:err}, we see that $E_\kappa$ decreases with $\kappa$, although its order of magnitude remains non-negligible. Hence, the diffusive model with fat tails may approach, in a sense, the non-diffusive one with slim tails, but visible differences remain between them as a consequence of the fact that the diffusive model describes a jackpot game which is not completely equivalent to the real one caught by the non-diffusive model.

\section{Conclusions}
\label{sec:conclusions}
In this paper, we introduced and discussed kinetic models of online jackpot games, i.e. lottery-type games which occupy a big portion of the web gambling market. Unlike the classical kinetic theory of rarefied gases, where binary collisions are dominant, in this case the game is characterised by simultaneous interactions among a large number $N\gg 1$ of gamblers, which leads to a highly non-linear Boltzmann-type equation for the evolution of the density of the gambler's winnings. When participating in repeated rounds of the jackpot game, the gamblers continuously refill the number of tickets available to play and, at the same time, their winnings undergo a percentage cut operated by the site which administers the game. Hence, through the study of the evolution of the mean number of tickets and of its variance, one realises that the solution of the model should approach in time a non-trivial steady state describing the equilibrium distribution of the gambler's winnings.

In the limit $N\to\infty$, we showed that the multiple-interaction kinetic model can be suitably linearised, so as to get access to analytical information about the large time trend of its solution. We proposed two different linearisations, which, while apparently both consistent with the original non-linear model, exhibit marked differences for large times. The solution to the linear model presented in Section~\ref{linea1} converges towards a steady state with all moments bounded. In some cases, such a steady state can be written explicit in the form of a gamma probability density function. Conversely, the solution to the linear model considered in Section~\ref{fat} converges towards a steady state in the form of an inverse gamma probability density function, hence with Pareto-type fat tails. We explained the different trend of the second model as a consequence of a too strong loss of correlation among the gamblers, which is instead present in the original non-linear multiple-interaction model and also in its linear approximation proposed in Section~\ref{linea1}. Numerical results showed indeed that the solution to this linear model is in perfect agreement with that to the full non-linear kinetic model.

The main conclusion which can be drawn from the present analysis is that the \textit{wealth economy} of a multi-agent system in which the trading activity relies on the rules of the jackpot game does not lead to a stationary distribution exhibiting Pareto-type fat tails, as it happens instead in a real economy. Unlike the real trading economy, where the small richest part of the population owns a relevant percentage of the total wealth, in the economy of the jackpot game the class of rich people is still very small but it does not own a consistent percentage of the total wealth (measured in terms of tickets played and won in time). In other words, it is exceptional to become rich by just playing the jackpot game and, in such a case, it is further exceptional to become very rich.

A non-secondary conclusion of the present analysis is that the rules of the jackpot game imply a strong correlation among the gamblers participating in the game. Indeed, in each round of the game there is just one gambler who wins, while all the other gamblers lose. Any approximation of the full non-linear model needs to take into account this aspect. This is clearly in contrast with a real trading economy, where the agents may instead take advantage simultaneously of their trading activity. 

\section*{Acknowledgements}
This research was partially supported by the Italian Ministry of Education, University and Research (MIUR) through the ``Dipartimenti di Eccellenza'' Programme (2018-2022) -- Department of Mathematics ``F. Casorati'', University of Pavia and Department of Mathematical Sciences ``G. L. Lagrange'', Politecnico di Torino (CUP: E11G18000350001) and through the PRIN 2017 project (No. 2017KKJP4X) ``Innovative numerical methods for evolutionary partial differential equations and applications''.

This work is also part of the activities of the Starting Grant ``Attracting Excellent Professors'' funded by ``Compagnia di San Paolo'' (Torino) and promoted by Politecnico di Torino.

All the authors are members of GNFM (Gruppo Nazionale per la Fisica Matematica) of INdAM (Istituto Nazionale di Alta Matematica), Italy.


\end{document}